\documentclass{aa}  

\usepackage{graphicx}
\usepackage{txfonts}
\usepackage{lipsum}
\usepackage{subcaption}
\usepackage{multirow}
\usepackage{lscape}             

\usepackage{pdflscape}
\usepackage{placeins}           
                                
\begin{document}

   \title{Cepheids with giant companions}
   \subtitle{III. Evolutionary modeling of nine binary double Cepheids from the Milky Way and Magellanic Clouds}

   \author{F. Espinoza-Arancibia\inst{1}
        \and B. Pilecki\inst{1} \and M. Catelan\inst{2} \and V. Hocdé\inst{1}
        \and I. B. Thompson\inst{3} \and W. Gieren\inst{4}}

   \institute{$^1$Nicolaus Copernicus Astronomical Center, Polish Academy of Sciences, Bartycka 18, 00-716 Warsaw, Poland\\
   $^2$Instituto de Astrofísica, Pontificia Universidad Católica de Chile,
Av. Vicuña Mackenna 4860, 7820436 Macul, Santiago, Chile\\
$^3$Carnegie Observatories, 813 Santa Barbara Street, Pasadena, CA 91101-1292, USA\\
$^4$Universidad de Concepción, Departamento Astronomía, Casilla 160-C, Concepción, Chile\\
             \email{fespinoza@camk.edu.pl}
            }


  \abstract
   {Binary double (BIND) Cepheids are systems comprising two Cepheid components. This feature provides important constraints that allow us to reveal the origin of Cepheids, trace their evolution, and test pulsation theory. 
   Ten BIND Cepheids are now known, with only one having its parameters determined. In five systems, the difference between the pulsational periods of Cepheid components is unusually high.}
   {We aim to estimate the physical parameters of the components of nine BIND Cepheids newly identified in the Magellanic Clouds and the Milky Way, investigate their evolutionary configurations, and formation scenarios. We also expand the parameter space of characterized individual Cepheids in mass, radius, period, and metallicity.}
   {To achieve these goals, we extended the recently introduced $q$-PED method to BIND Cepheids, combining observational constraints with theoretical pulsation and evolutionary models. With the pending determination of the spectroscopic mass ratio ($q_s$), we used the pulsation periods of both components in its place. We considered all consistent configurations (first-crossing, blue-loop, and mixed) as viable solutions. Probabilistic and observational constraints, including spectroscopic mass ratios for two systems, were then used to discriminate between them.}
   {We obtained new $q$-PED estimates of mass, radius, temperature, luminosity, and age for 18 Cepheids with previously unknown physical parameters. For one Galactic system, the spectroscopic mass ratio $q_s=0.84\pm0.04$ indicates a first-crossing plus a blue-loop Cepheid solution. This mass ratio, along with the predicted mass ratios lower than unity for two other systems, suggests past binary interactions, most likely a merger origin for one of the components. In addition, we derive a new period--mass--radius relation and fit a mass--luminosity relation covering the mass range $2.3-4.6$ M$_\odot$ that clearly separates between different crossings of the instability strip.}
   {This work provides the first mass estimates for Cepheids in the SMC, extending the lower Cepheid mass limit down to 2.3 M$_\odot$. For the benchmark eclipsing Cepheid OGLE-LMC-CEP-1718, pulsationally driven mass loss may account for its unexpected, reversed mass ratio. Binary interactions in the past evolution of Cepheids may be common, affecting up to 40\% of our systems with two clear cases and two more if blue loop Cepheids are preferred. Improved spectroscopic mass ratios from ongoing observational campaigns will gradually clarify the evolutionary history for all presented BIND Cepheids.}

   \maketitle
   \nolinenumbers
\section{Introduction}

Arguably, one of the most important classes of pulsating stars is that of classical Cepheids (hereinafter referred to simply as Cepheids). These stars exhibit a well-known and close relationship between their pulsation period and luminosity \citep{Leavitt1912}, which places them as an important step in determining distances within our galaxy and the Local Group \citep[for a recent review, see][]{Bono2024}. Moreover, they are valuable objects for testing stellar evolution and pulsation models \citep[see, e.g.,][]{Hocde2024, Stuck2025, Deka2025, Espinoza2025}.

Cepheids are frequently found in binary or multiple systems. Their binary fraction is estimated to exceed 80\% \citep{Kervella2019}, suggesting that their evolutionary paths and pulsation properties can be significantly influenced by binary interactions, such as tidal deformation, mass transfer, or mergers \citep[see, e.g.,][]{2018ApJ...868...30P, Espinoza-Arancibia2025}. At the same time, this high multiplicity offers valuable opportunities, as Cepheids in binary systems allow direct and accurate determinations of fundamental stellar parameters, most notably their masses, which play a central role in stellar evolution theory. Currently, six Cepheids in double-lined eclipsing binary systems in the Large Magellanic Cloud (LMC) have been studied \citep{Pilecki2018}, obtaining precisions of 0.5-2.5\% for the most relevant masses and radii \citep[for a short review, see][and references therein]{Pilecki2025review}. In the Milky Way (MW), masses have been determined for seven Cepheids in optical single-lined binaries \citep{Evans2018, Gallenne2018, Gallenne2019, Evans2024b, Evans2024a, Gallenne2025}. These measurements are obtained in two ways. The first approach uses the mass ratio derived from high-resolution ultraviolet spectroscopy (where both components can be detected) and the companion's mass from the spectral energy distribution. This method has low accuracy and a precision of 10-20\%. The second approach is available for a smaller number of binary Cepheids and uses interferometric observations to constrain the orbital semi-major axis and inclination. Along with the distance to the system, these parameters allow for robust mass determinations with a precision of 1.1-5.5\%. Unfortunately, as of now, there have been no mass determinations of Cepheids in the Small Magellanic Cloud (SMC).

Overall, these observational data are still limited. The aforementioned Cepheid mass determinations range between 3.6-5 M$_\odot$, with only one value higher than 6 M$_\odot$, but with a high uncertainty of 13$\%$. Theoretical studies predict that Cepheids span a mass range between 3-11 M$_\odot$ \citep[see, e.g.,][]{Bono2000,Anderson2016}. Therefore, the mass--luminosity (ML) relation is poorly constrained \citep{Anderson2016}. Furthermore, at the LMC metallicity, the blue loops (BLs) predicted by evolutionary tracks are too short to account for the presence of low-mass short-period Cepheids \citep{Espinoza-Arancibia2024}. At the SMC metallicity, the extent of the BLs is also inconsistent with observations \citep{Espinoza2025}. To address these issues, Cepheid masses over a broader range are essential. In particular, Cepheids in spectroscopic double-lined (SB2) binaries are the best candidates for measuring their mass, since we do not expect to find many more eclipsing binaries. 

In the first paper of this series \citep[Paper I;][]{Pilecki2021}, we introduced a new approach for identifying Cepheids in SB2 systems, in which Cepheids that are overluminous for their periods, exhibit similar or redder colors, and show reduced pulsation amplitudes that are interpreted as having evolved giant companions. In Paper II \citep{Pilecki2024a}, we reported the spectroscopic detection of binarity in nine double Cepheids with orbital periods ranging from 2 to 18 years. This study increased the number of known binary double (BIND) Cepheids from one to ten, and tripled the total number of confirmed SB2 Cepheids. In this paper, we aim to estimate the physical parameters for the nine BIND Cepheids using, with a slight modification, a theoretical modeling method, introduced in \cite{Espinoza-Arancibia2025}, that incorporates empirical constraints. Since direct dynamical masses are unavailable for these systems, deriving physical parameters through a combined evolutionary and pulsational modeling approach can substantially expand the range of sampled Cepheid masses. These inferred masses have the potential to also allow us to establish new relations involving this fundamental quantity across three galaxies with different metallicities.

The outline of this paper is as follows. In Section~\ref{sec:sample}, we present the sample of binary double Cepheids studied in this work. In Section~\ref{sec:qped}, we describe the extended $q$-PED method adapted for BIND Cepheids, including the grids of evolutionary and pulsation models and the multiband distance-fitting procedure. In Section~\ref{sec:results}, we present the results of applying the method to the benchmark eclipsing system LMC-CEP-1718 and to the nine newly identified BIND Cepheids. In Section~\ref{sec:disc}, we discuss the minimum intrinsic mass ratio expected for non-interacting BIND systems, the most likely origin and evolutionary status of each system, and the derived period--mass--radius (PMR) and ML relations. Finally, in Section~\ref{sec:conclusions}, we summarize our main conclusions.

\section{Sample of binary double Cepheids}\label{sec:sample}

Table~\ref{table:1} summarizes the modes and periods of the newly identified BIND Cepheids from Paper II, including the eclipsing Cepheid LMC-CEP-1718, studied previously in \citet{Pilecki2018}.  Among the new systems, two are located in the MW, five in the SMC, and two in the LMC. Therefore, the sample spans a significant range of metallicities. Furthermore, the sample includes all possible combinations of fundamental (F) and first-overtone (1O) modes, spanning a wide range of period ratios. The primary component (component 1) is defined as the Cepheid with a longer period (fundamentalized for the overtone mode) and the secondary (component 2) as the one with a shorter period in the system. Accordingly, the period and mass ratios are defined as $P_2/P_1$ and $M_2/M_1$, respectively.

Typically, the components of binary systems originate from the same interstellar cloud, which means they share a similar age and chemical composition. Since Cepheids are usually found on BLs, when two are found in the same binary system, they should have comparable masses and radii, constrained by the limits of the instability strip (IS) and the characteristics of the loop. Additionally, the pulsation period depends on mass and radius \citep[see, e.g.,][]{Catelan2015}. Therefore, a binary double Cepheid should present similar periods. Contrary to our expectations, more than half of our sample exhibits period ratios below 0.8 (see Table~\ref{table:1}). The likely explanation of this phenomenon is the combination of a first-crossing (1C) Cepheid (evolving along the subgiant branch) with a more typical BL Cepheid. Alternatively, binary interactions such as mass transfer or stellar mergers could produce Cepheids of different masses that nevertheless share the same age \citep[see e.g.,][]{Pilecki2022, Espinoza-Arancibia2025}. As both 1C and merger-origin Cepheids are extremely interesting, rare, and generally hard to identify, the analysis of this sample will yield valuable results regardless of which scenario ultimately accounts for the large difference in pulsation periods between the components of BIND Cepheids.

\begin{table*}[h!]
\caption{Pulsation modes and periods of the binary double Cepheid sample}
\label{table:1}
\centering 
\begin{tabular}{c c c c c c c}
\hline\hline
OGLE ID & Modes & $P_1$ [days] & $P^{\rm F}_1$ [days]& $P_2$ [days] & $P^{\rm F}_2$ [days] & $P^{\rm F}_2$/$P^{\rm F}_1$\\
\hline
   BLG-CEP-067 & 1O $+$ 1O & $2.610721$ & $3.827$ & $1.692381$ & $2.444$ & $\textbf{0.639}$\\ 
   GD-CEP-0291 & F $+$ F & $3.667693$ & $3.668$ & $3.398977$ & $3.399$ & $0.927$\\
   LMC-CEP-0571 & F $+$ 1O & $3.079937$ & $3.080$ & $2.100885$ & $3.057$ & $0.992$\\
   LMC-CEP-0835 & F $+$ F & $4.562781$ & $4.563$ & $2.750956$ & $2.751$ & $\textbf{0.603}$\\
   LMC-CEP-1718 & 1O $+$ 1O & $2.480909$ & $3.649$ & $1.963683$ & $2.869$ & $\textbf{0.786}$\\
   SMC-CEP-1526 & F $+$ F & $1.804311$ & $1.804$ & $1.290234$ & $1.290$ & $\textbf{0.715}$\\
   SMC-CEP-2699 & 1O $+$ F & $2.562225$ & $3.772$ & $2.117341$ & $2.117$ & $\textbf{0.561}$\\
   SMC-CEP-2893 & F $+$ F & $1.321549$ & $1.322$ & $1.135859$ & $1.136$ & $0.860$\\
   SMC-CEP-3115 & F $+$ F & $1.251945$ & $1.252$ & $1.159784$ & $1.160$ & $0.926$\\
   SMC-CEP-3674 & F $+$ 1O & $2.896089$ & $2.896$ & $1.827785$ & $2.665$ & $0.920$\\
\hline                    
\end{tabular}
\tablefoot{Period ($P_i$) and fundamentalized period ($P^{\rm F}_i$) for each component ($i=1,2$) of the binary double Cepheids. Periods were gathered from the Optical Gravitational Lensing Experiment Catalog (OGLE-IV) catalog \citep{Soszynski2017}. Fundamentalized periods were obtained using the relations provided in \citet{Pilecki2024}. Period ratios lower than 0.8 are shown in boldface.}
\end{table*}

\section{The $q$-PED method on binary double Cepheids}\label{sec:qped}

The $q$-PED method was recently introduced in \cite{Espinoza-Arancibia2025} to determine the physical parameters of a likely merger-origin binary Cepheid (OGLE-LMC-CEP-1347). This method combines theory and observations, using the measured mass ratio ($ q = M_2/M_1$), pulsation (P), and evolutionary (E) models, along with the known distance (D) and multiband photometry, to constrain the physical parameters of the components of a binary Cepheid. The $q$-PED method uses evolutionary models computed with the version r22.11.1 of the modules for experiments in stellar astrophysics \citep[MESA,][]{Paxton2019, Jermyn2023} code. On the other hand, pulsation models are computed with the radial stellar pulsation (RSP) functionality of MESA \citep{Paxton2019}.

RSP is a 1D-hydrodynamical code, introduced by \citet{Smolec2008}, that generates a starting convective model of the envelope, performs a linear analysis that yields periods and growth-rates, and later performs non-linear calculations. We restricted our RSP pulsation models to the linear regime and adopted the set D of convective parameters from \citet{Paxton2019}. It is worth noting that no single set of convective parameters provides a satisfactory description of Cepheid pulsation across the entire IS \citep[see, e.g.,][]{Deka2026}. Nonetheless, set D has been shown to reproduce the empirical boundaries of the LMC IS well \citep{Espinoza-Arancibia2024}. Furthermore, since pulsation periods are largely insensitive to the choice of convective parameters \citep{Paxton2019}, we do not expect our results to depend significantly on this choice.

In short, the $q$-PED method consists of three main parts. First, we calculate an initial grid of evolutionary tracks for the components of the binary system. Second, we compute linear pulsational models for every point along the Cepheid’s evolutionary track within the IS, and we select those pulsating with the same observed period value as the Cepheid in the system. Finally, we use the known distance to the object and the measured mass ratio as constraints to identify valid system configurations for the Cepheid pulsational models and the positions of the companion on the corresponding evolutionary tracks.

In the case of BIND Cepheids, both components of the system lie within the IS, which drastically reduces the parameter space and allows their pulsations to be modeled. This useful feature allows us to treat the mass ratio $q$ as a free parameter. Therefore, using the $q$-PED method, we can also obtain preliminary parameters of BIND Cepheids only using the Cepheids' periods, without prior information about the mass ratio. However, when the mass ratio is measured, we can establish tighter constraints on the system configuration and the parameters of its components.

The modified $q$-PED method for BIND Cepheids works as follows: 
\begin{enumerate}

\item We compute a dense grid of evolutionary tracks with metallicities consistent with the MW ($Z=0.009$, 0.010, and 0.011), LMC ($Z=0.006$, 0.008, and 0.009), SMC ($Z=0.001$, 0.002, and 0.003). We initially consider a mass range of 3 to 4 $M_{\sun}$ for the three galaxies. This range was iteratively extended to cover all valid models of our sample. For the LMC, we considered a final mass range of 3 to 5 $M_{\sun}$. For the SMC models, we considered a mass range of 2 to 4 $M_{\sun}$, and for the MW models, a mass range of 3 to 5.5 $M_{\sun}$. In all cases, the mass resolution was 0.01  $M_{\sun}$.

\item We compute linear pulsating models using RSP for all the points of the evolutionary tracks that lie within the LMC and SMC empirical IS edges \citep{Espinoza-Arancibia2024, Espinoza2025}. To simplify the analysis, we considered the empirical wedge-shaped IS instead of an IS with a break. The differences in $(V-I)$ color between the two formulations in the magnitude range of our LMC and SMC samples are lower than $0.05$ mag. This is not significant compared to the current precision of our solutions. Once we can measure precise mass ratios to use as constraints, we will perform a more precise analysis that accounts for breaks in the IS. For models with MW metallicity, we used the LMC empirical IS borders to approximate our galaxy's IS\footnote{This is not a perfect approximation, but we still consider it superior to introducing any theoretical IS, which show significant shifts with respect to the empirical ones \citep{Espinoza2025}. Once the empirical IS for MW is available, we will use it in our future work.} From the models inside the IS, we then select those RSP models pulsating with the same period values as the Cepheid components, within a margin of $5\%$\footnote{As shown in \citet{Espinoza-Arancibia2025}, adopting a larger margin of 20\% produces results that remain consistent with those obtained using the smaller margin. The only notable difference is found in the radius, which exhibits an increase of 2\% when the larger margin is applied to the periods.}. The use of this margin yielded pulsation models consistent with the observed properties of the system OGLE-LMC-CEP-1812, and produced physical parameters in agreement with an empirical PMR relation \citep{Espinoza-Arancibia2025}. We then coadd the fluxes from the RSP models for both Cepheids and compute the binary system's magnitude in different photometric bands.

\item We constrained the system magnitudes to reproduce the known distance moduli of the BIND Cepheids host galaxies. This step was performed using the multiband method, in which we fit the relationship
\begin{equation}
    (m-M)_0 = (m-M)_\lambda - E_{B-V}R_\lambda ,
\end{equation}
on each model. We assumed the values of the total-to-selective absorption from \citet{Breuval2022}, which are based on the reddening law by \citet{Fitzpatrick1999}\footnote{Note that this method is not very sensitive to the exact reddening law applied \citep{Wielgorski2017}.}, and the reddened distance moduli $(m-M)_\lambda$ from multi-band photometry and each of our models. The resulting slope and intercept of the linear fit provide an estimate of the true distance modulus $(m-M)_0$ and the reddening $E_{B-V}$. We then limited the obtained $(m-M)_0$ to the empirical values of the host galaxies of the systems. In particular, we considered the LMC distance modulus determined by \citet{Pietrzynski2019} of $(m-M)_0=18.477 \pm 0.030\rm~mag$, and the SMC distance modulus by \citet{Graczyk2020} of $(m-M)_0=18.977\pm0.044\rm~mag$. In the case of the systems in the MW, we adopted their distances calculated in \citet{Pilecki2022}.
\end{enumerate}

The considered multi-band photometry for each system, and the derived reddening from the multiband method are presented in Appendix~\ref{table:A1-photometry}. After selecting the valid models that meet all established constraints, we calculate an unweighted average of the parameter values, which form the system's final inferred characteristics. The reported uncertainties represent the standard deviation of the accepted models.

\section{Results}\label{sec:results}

\subsection{LMC-CEP-1718}

LMC-CEP-1718 was the first confirmed and well-studied eclipsing BIND Cepheid. Its precise orbital and physical parameters were determined from spectroscopic observations \citep{Pilecki2018} and light-curve modeling. For this reason, this system serves as a benchmark for evaluating the accuracy of the $q$-PED method in recovering the component parameters.

We modeled the system using three configurations of the $q$-PED method. In the first two configurations, we treated the predicted mass ratio ($q_{\rm p}$) as a free parameter, allowing it to vary between 0.5 and 1.2, while the ages of the two components were treated as free parameters in the first configuration and as fixed parameters in the second. Conversely, the third configuration assumed the same age for both components and a fixed mass ratio, constrained by the empirical value $q_{\rm s}$ determined from spectroscopy. The obtained parameters are shown in Table \ref{tab:LMC-CEP-1718-comparison}.

In the first case, most derived parameters agreed with the observed values within uncertainties, with the radii of both components differing by around 1\%. However, we found a mass ratio greater than unity, indicating that our secondary component is more massive than the primary, which contradicts the empirical results. In the second configuration, we introduced an additional constraint requiring the components' ages to be approximately equal (to within 1\%). The resulting age is $150 \pm 10\rm~Myr$. This constrained method yields results with better overall agreement with observations (2.1$\%$ on average, compared to 3.5$\%$ in the first case), although the mass ratio remains inverted relative to the observed values. In the last configuration, we constrained the mass ratio to the observed value to within the uncertainties, along with the components' ages. In this case, the age of both components yielded $140 \pm 10\rm~Myr$. Overall, the results are comparable to those in the previous case (with an average agreement of 2.5$\%$), except for the primary mass, which increased, leading to better agreement with observations. However, for our BIND Cepheids sample, the empirical mass ratios are still preliminary, and for the sake of this work, we treat the results obtained with a constrained age and an unconstrained mass ratio as indicative of the accuracy and precision of our results. Nevertheless, with future improvements in determining the mass ratios, we will be able to refine our results considerably. Similar to what was found in \citet{Espinoza-Arancibia2025}, we expect that systematic errors in the $q$-PED should not exceed 4\% regardless of the applied constraints. On the other hand, LMC-CEP-1718 is the only (out of 5) eclipsing binary that exhibits a less-massive secondary appearing more advanced evolutionarily. Consequently, we treat this and the corresponding systematic effect on the parameters as the worst-case scenario that may affect only a small subset of the analyzed systems.

\begin{figure}
    \centering
    \includegraphics[width=\linewidth]{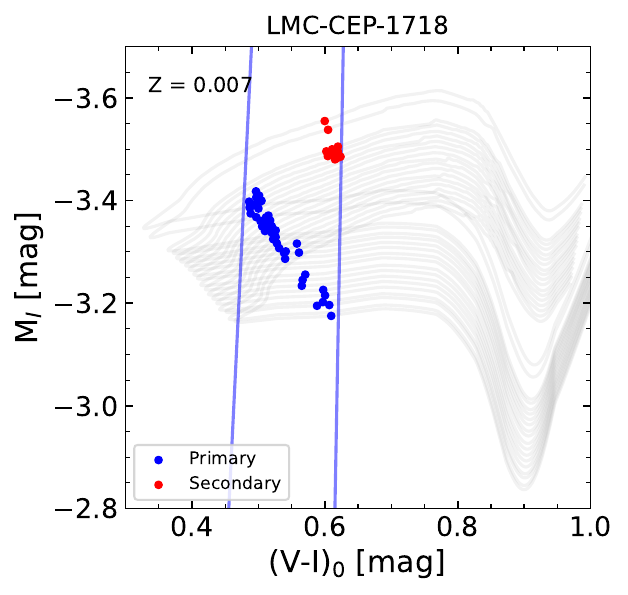}
    \caption{Color-magnitude diagram showing the obtained valid models of LMC-CEP-1718, for one value of metallicity. In gray lines, we display the evolutionary tracks from which the models were taken. The blue and red points represent the models for the primary and secondary Cepheids, respectively. The blue lines are the empirical first-overtone instability strip edges from \citet{Espinoza-Arancibia2024}.}
    \label{fig:LMC-CEP-1718-CMD}
\end{figure}

In Fig.~\ref{fig:LMC-CEP-1718-CMD}, we present the final positions of the primary and secondary components in the color-magnitude diagram (CMD) obtained using the $q$-PED method. Since we fixed the system's age and the secondary component is more massive, its position in the CMD indicates it is also more evolved and undergoing the third crossing of the IS, as expected. Interestingly, the latter observation was also reported in \citet{Pilecki2018}, although their spectroscopic mass ratio indicates that the secondary component is less massive than the primary. Given that we do not account for binary interactions in our modeling, the mass discrepancies between our results and the empirical values could indicate that LMC-CEP-1718 experienced a minor mass transfer event. The total mass transferred would be about 0.06 $M_{\sun}$. The mechanism responsible for this hypothetical mass transfer, however, remains uncertain. The highest chance of occurrence would be during the red giant branch (RGB) phase, but evolutionary models show that stars of these masses would be far from overfilling their Roche lobes even at their maximum radii at the RGB phase peak. They can reach radii of up to 64 R$_{\odot}$, while their Roche lobe radii are about 180 R$_{\odot}$. Therefore, mass transfer due to Roche-lobe overflow is unlikely. Another possibility may be pulsation-driven mass loss, which would have occurred more recently. The initial primary would lose more mass during its second and third crossings than the initial secondary would during only its partial second crossing, leading to a small-scale mass reversal and a role interchange. \citet{Neilson2011} estimated that a Cepheid may lose 5-10\% of its mass due to this effect, so a few percent difference in the mass lost between more and less evolutionarily advanced components is perfectly possible. At the moment, we treat it as the most likely explanation for the small anomaly in this system's mass ratio. However, an alternative solution with different evolutionary paths and rates of both components due to different initial rotation cannot be fully excluded.
 
\begin{table*}
\caption{Results for the eclipsing binary Cepheid LMC-CEP-1718}
    \centering
    \begin{tabular}{cccccccccccccc}
    \hline\hline
OGLE ID& $M$ [M$_\odot$] & $T_{\rm eff}$ [K]& $\log L$ [L$_\odot$] & $\log g$ [cm/s$^2$]&$R$ [R$_\odot$] & $q$ & Comment\\\hline

LMC-CEP-1718A &$4.16(11)$ & $6192(112)$ & $3.01(4)$ & $2.17(1)$ & $28.17(55)$& 1.08(4) & This work - free age\\
              & \textbf{4.25(5)} & \textbf{6258(69)}  & \textbf{3.04(2)} & \textbf{2.16(1)} & \textbf{28.51(43)}& \textbf{1.02(2)} & \textbf{This work - fixed age}\\
              & $4.34(6)$ & $6306(44)$ & $3.07(1)$ & $2.16(1)$ & $29.03(29)$ & 0.989(3)& This work - fixed $q$\\
              & 4.27(4) &  6310(150) & 3.04(6) & 2.14(4)  & 27.8(12) & 0.988(5)& \cite{Pilecki2018}\\\hline

LMC-CEP-1718B &$4.49(11)$ & $6145(100)$ & $3.15(3)$ & $2.05(1)$ & $33.46(55)$ & 1.08(4) & This work - free age \\
              & \textbf{4.34(6)} &	 \textbf{6015(49)}  &  \textbf{3.11(2)} &	\textbf{2.04(1)} & \textbf{33.25(49)} & \textbf{1.02(2)} & \textbf{This work - fixed age}\\
              &$4.29(6)$ & $6023(69)$ & $3.09(2)$ & $2.05(1)$ & $32.63(42)$ & 0.989(3)& This work - fixed $q$\\
              & 4.22(4) &  6270(160) & 3.18(6) & 2.02(3) & 33.1(13) & 0.988(5)& \cite{Pilecki2018}\\\hline
    \end{tabular}
    \label{tab:LMC-CEP-1718-comparison}
\tablefoot{Results of the modeling using the $q$-PED method, considering three different configurations: free age and mass ratio, fixed age and free mass ratio (marked in boldface), and fixed age and mass ratio. The empirical results of the system, from \citet{Pilecki2018}, are also displayed for comparison. The second configuration was adopted during the modeling of the subsequent BIND Cepheids.}
\end{table*}

\subsection{New BIND Cepheids sample}

\begin{figure*}
    \centering
    \includegraphics[width=\hsize]{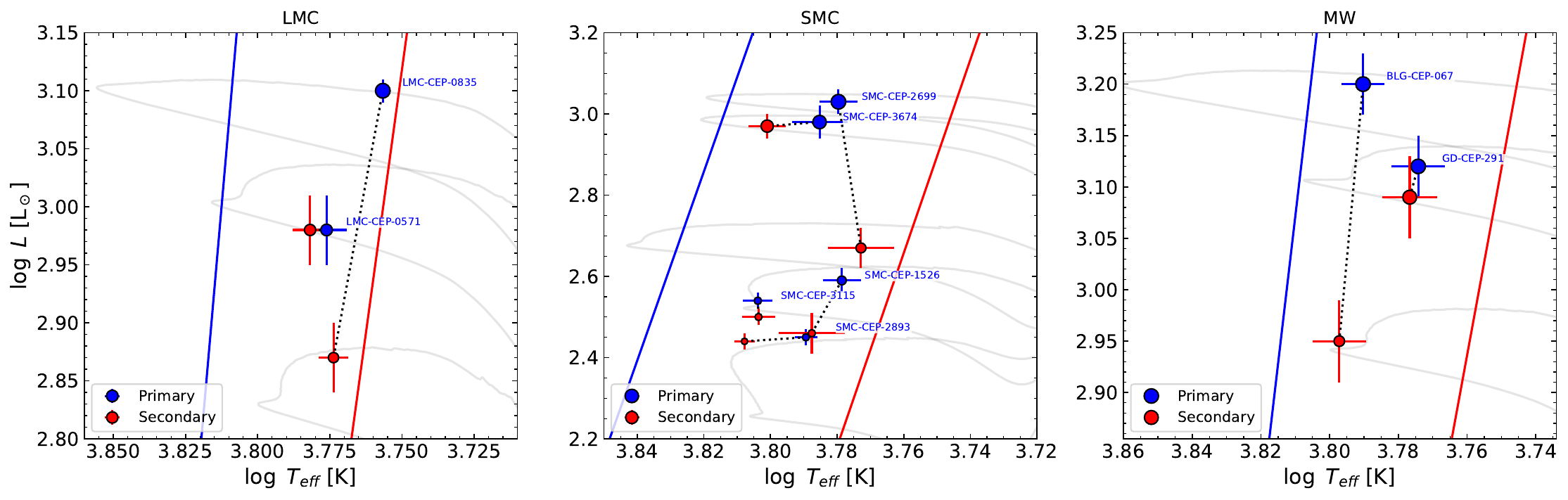}
    \caption{Hertzsprung–Russell diagrams showing the average BL solutions of the sample of BIND Cepheids. The blue and red circles represent the average models of the primary and secondary Cepheids, respectively. The dotted black lines connect both components of the system. Results for the LMC, SMC, and our galaxy are shown in the left, central, and right panels, respectively. In gray lines, we display the evolutionary tracks of the corresponding average mass of the systems' components. The blue and red lines correspond to the empirical instability strip edges from \citet{Espinoza-Arancibia2024,Espinoza2025}, considering both F and 1O pulsation modes.}
    \label{fig:Solutions_BL}
\end{figure*}

\begin{figure*}
    \centering
    \sidecaption
    \includegraphics[width=0.7\hsize]{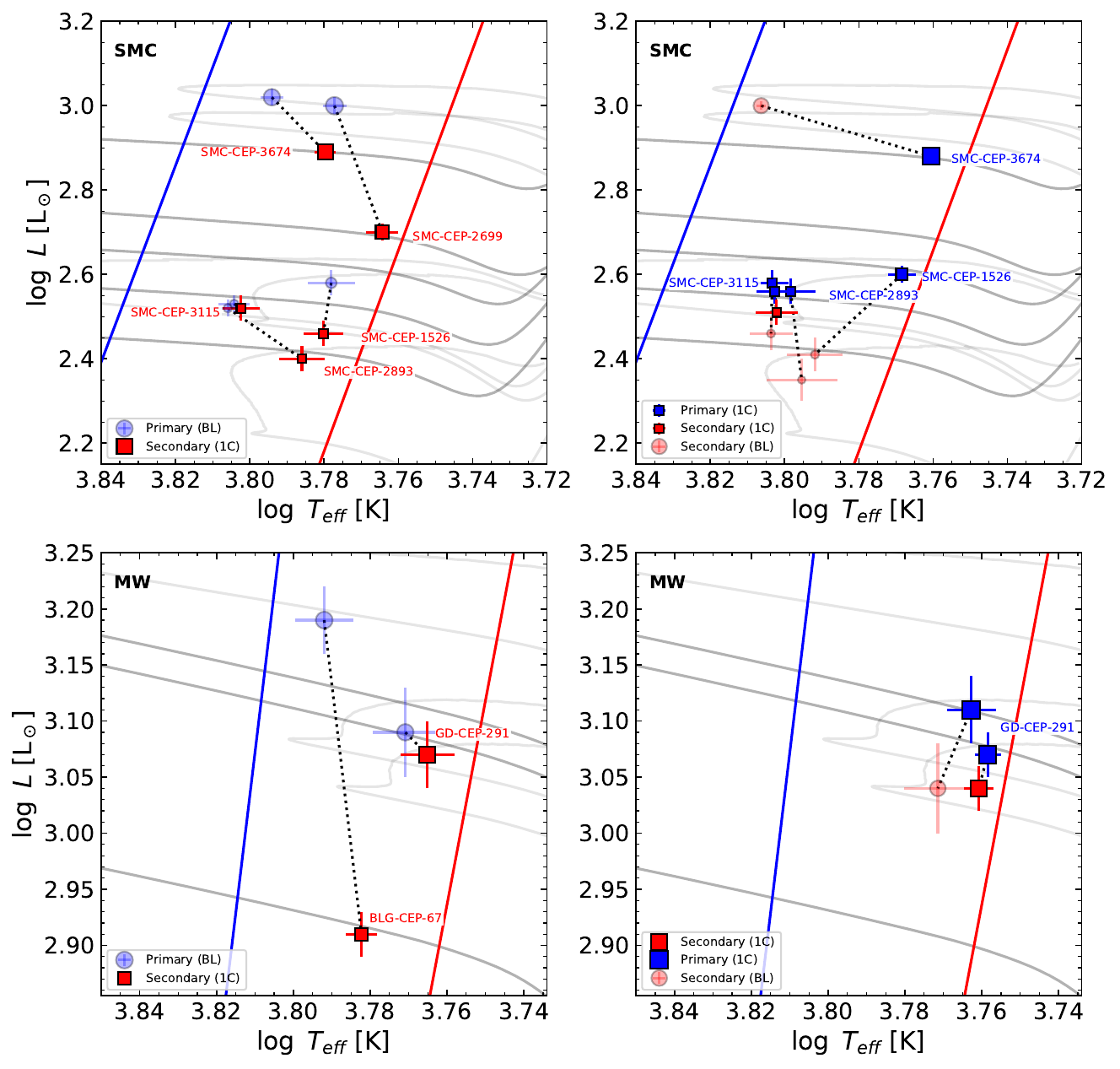}
        \caption{The same as Fig.~\ref{fig:Solutions_BL} for the average solutions that involved 1C Cepheids. Solutions for the SMC and MW are shown in the upper and lower panels, respectively. In addition, the left panels only display solutions featuring a BL primary and a 1C secondary. The right panels include solutions with two 1C components and a 1C primary with a BL secondary. The blue and red transparent circles represent the BL primary and secondary Cepheids, respectively. Conversely, the blue and red squares represent the 1C primary and secondary Cepheids, respectively.}
    \label{fig:Solutions_1C}
\end{figure*}

In the following section, we present the results obtained for the BIND Cepheids in our sample. It is important to note that in some systems, we identified multiple valid solutions. This is because a region within the IS can be populated by either BL Cepheids of a given mass or by 1C Cepheids of a higher mass. In general, we found all combinations of 1C and BL Cepheids as possible solutions. We separated them into distinct solutions based on their evolutionary status. In Fig.~\ref{fig:Solutions_BL}, we show the location in the Hertzsprung–Russell diagram (HRD) of the solutions presenting two BL Cepheids. In addition, Fig.~\ref{fig:Solutions_1C} shows the same for solutions involving 1C Cepheids for SMC and MW Cepheids. LMC Cepheids are not included, because this combination of IS crossings was not found for systems in that galaxy. In Appendix~\ref{fig:all_BL} and \ref{fig:CMDs_1st}, all valid pulsation models for the solutions involving BL and 1C are shown. Additionally, Appendix~\ref{fig:multiband} shows the multi-band fits obtained for the BL solutions.

Given the higher a priori probability of finding a Cepheid in the BL evolutionary phase, we adopt the configurations involving BL Cepheids as the preferred solutions, unless independent evidence indicates the presence of 1C Cepheids. Therefore, in Appendix~\ref{tab:results}, we present the average physical parameters of the solutions that are composed of two BL Cepheids. In the cases where the mass ratio was sufficiently close to unity, allowing for the presence of two coeval (same-age) Cepheids, we constrain their ages to be similar to within 1\%. This is based on the assumption that no significant past interaction occurred that would change their masses. In addition, in Appendix~\ref{tab:results_1st} we included the solutions that involve 1C Cepheids. We discuss the possible origins of the BIND systems and the most likely solutions in more detail in Section~\ref{subsec:origins}.

For the two Galactic systems, we found multiple valid configurations combining 1C and BL Cepheids. In both cases, the BL+BL solutions yield component masses in the range 3.95–4.63~$M_\odot$. For BLG-CEP-067, we found solutions (BL+BL and BL+1C) considering a metallicity value of $Z=0.006, 0.007,$ and $0.008$ ([Fe/H]$\sim-0.45$, $-0.38$, and $-0.33$, respectively), all of which fall within the confidence interval of the metallicity gradient of the Galactic disk \citep[see, e.g.,][]{Spina2022}. The BL+BL configuration yielded a mass ratio of 0.87 $\pm$ 0.03 and an age difference of $50$ Myr. For GD-CEP-291, we extended the grid of evolutionary tracks with $Z=0.009, 0.01, 0.011$ up to 5.5 $M_{\sun}$ to account for all possible solutions. We determined four distinct configurations (BL+BL, BL+1C, 1C+BL, 1C+1C), with masses ranging from $\sim$4.43 to $\sim$5.33 M$_\odot$ depending on the configuration. The BL+BL solution yields nearly identical masses ($\sim$4.60 M$_{\odot}$) and ages ($\sim$130 Myr) for both components. The remaining solutions involve age differences of $\sim~$40 Myr and mass ratios ranging from 0.83 to 1.15.

For the two LMC systems, we did not find 1C configurations; only BL+BL solutions were found. LMC-CEP-0571 yielded a single configuration with nearly identical component masses ($\sim$4.06 M$_\odot$) and comparable ages, consistent with a mass ratio close to unity. For LMC-CEP-0835, solutions were found only at lower metallicities (Z = 0.005–0.006), yielding a mass ratio of 0.88 $\pm$ 0.02 and an age difference of 50 Myr. 

The five SMC systems span the broadest range of configurations found in this study. SMC-CEP-3115 admits all possible solutions. SMC-CEP-3674, SMC-CEP-2893, and SMC-CEP-1526 yield three solutions each, while we found two solutions for SMC-CEP-2699. For the BL solutions, masses range from 2.32 to 3.58 M$_\odot$, extending the parameter space of characterized Cepheids to significantly lower masses than previously available. The mass ratios of the BL solutions span 0.80-1.04, while the alternative configurations involving 1C Cepheids yield mass ratios of 0.69-1.16, with age differences of 60-370 Myr. 

\section{Discussion}\label{sec:disc}

\subsection{Minimum mass ratio due to different IS crossings}\label{subsec:massratio}

Appendix~\ref{tab:results} and Appendix~\ref{tab:results_1st} show that the mass ratios obtained range from $0.69$ to $1.15$. In particular, considering only the solutions that presented a 1C primary, 4 out of 5 have mass ratios below $0.9$. Many external processes could change this value, such as mass transfer events or mergers \citep[see, e.g.,][]{Marchant2024}. On the other hand, we can expect an intrinsic variation of $q_{\rm p}$ from components of a binary that were born at the same time, with slightly different masses. For a BIND Cepheid, the system should have a mass ratio that allows both components to evolve at rates that place them within the IS. For a given age, the configuration that results in the smallest mass ratio would consist of a primary, more massive component located near the red edge of the third IS crossing. Meanwhile, the secondary component should be entering the IS for the first time. Additionally, we considered another case: a primary component at the end of the third IS crossing and a secondary component entering the second IS crossing. To study the possible mass ratios obtained from the previous configurations, we constructed a grid of isochrones from our evolutionary tracks. We used an equivalent evolutionary point framework \citep[see, e.g.,][]{Dotter2016}, in which stellar tracks were resampled, ensuring that phases characterized by rapid structural evolution (e.g., the Hertzsprung gap) were sampled with higher resolution than slowly evolving phases. For a given age, we located the intersection between the isochrones and the empirical IS of the LMC and SMC \citep{Espinoza-Arancibia2024, Espinoza2025}. Then, we calculated the mass ratio of the two configurations mentioned above. For the LMC, we used isochrones with a metallicity of $Z=0.007$ and a $\log \rm (age)$ range of $8.04$ to $8.27$. For the SMC, we considered a metallicity of $Z=0.002$ and $\log \rm (age)$ between $8.0$ and $8.63$. Finally, for the Milky Way, we used isochrones with $Z=0.01$ and $\log \rm (age)$ between $7.95$ and $8.15$. The upper limits of the adopted age ranges were selected so that the BLs of the oldest isochrones crossed the IS. On the other hand, the lower bounds are limited by the higher masses of our evolutionary tracks.

\begin{figure}
    \centering
    \includegraphics[width=\linewidth]{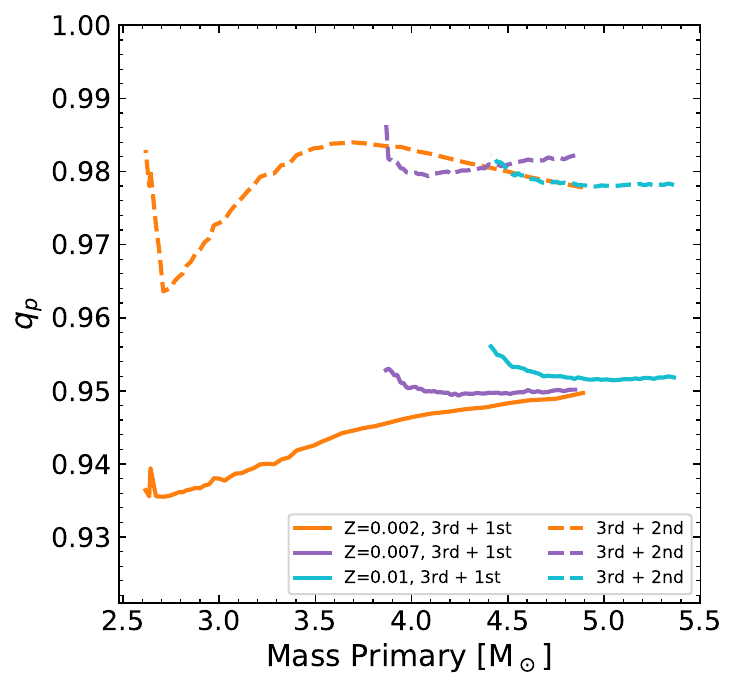}
    \caption{Predicted mass ratio as a function of the primary component mass. Two types of configuration are presented; the solid lines indicate the mass ratio between Cepheids in the third and first IS crossings, while the dashed lines represent the mass ratio between the third and second IS crossings. Three metallicities are also included: $Z=0.002$ (orange lines), $Z=0.007$ (purple lines), and $Z=0.01$ (cyan lines).}
    \label{fig:m1m2_q_age}
\end{figure}

In Appendix.~\ref{fig:Iso-IS}, we present the isochrones constructed, together with the IS edges. Fig.~\ref{fig:m1m2_q_age} shows the predicted mass ratios as a function of the mass of the primary component for metallicities representative of each galaxy. For both crossing configurations considered, the mass ratio generally starts with an abrupt decrease, followed by a steady increase as the primary mass increases. The starting high value of $q_{\rm p}$ arises from isochrones with the shortest BLs that briefly enter the IS. For the case of a first- and third-crossing Cepheids, the limiting $q_{\rm p}$ converges at roughly the same value of around 0.95 for the highest-mass primaries, with a slightly higher value for $Z=0.01$. Probably because of the extension to lower masses, for $Z=0.002$, the minimum mass ratio reaches significantly lower values, down to 0.936 for a primary mass around 2.7 M$_\odot$.

On the other hand, when we considered Cepheids in the third and second IS crossings, the mass ratio converged around $0.98$ or slightly above. As in the previously described configuration, for $Z=0.002$ the mass ratio can be significantly lower, reaching $0.964$ for low-mass primaries. These values represent observationally constrained model-based estimates of the minimum mass ratio for a BIND Cepheid system that evolves without any mass exchange between its components. Based on these single-star evolutionary models, lower values of $q_{\rm p}$ would require that the system's components have significantly different masses and apparent ages, suggesting a prior mass-exchange event.

\subsection{Possible origins and current evolutionary status of BIND Cepheids}\label{subsec:origins}

Our results, as presented in Appendix~\ref{tab:results} and \ref{tab:results_1st}, show that multiple evolutionary configurations are possible for each system. These different solutions have important implications for understanding the origin of BIND Cepheids. The use of empirical constraints to identify the most probable configuration is therefore essential.

In eight out of nine systems, at least one solution yields a mass ratio close to unity and nearly equal component ages. In such cases, the stars are likely evolving independently, without any significant past interactions. The alternative solutions exhibit mass ratios that are significantly different from unity, suggesting evolutionary histories influenced by binary interactions.

As discussed in \cite{Espinoza-Arancibia2025}, low mass ratios combined with age differences may point to mass transfer, stellar capture, or merger events. For example, \cite{Neilson2015} showed that systems with orbital periods shorter than about 240 days are expected to undergo Roche-lobe overflow before the star reaches the Cepheid phase. However, the shortest orbital period in our sample (for SMC-CEP-2893) is about three times longer (about 760 days), which yields a separation too large for mass transfer between the components, even at their maximum sizes on the red giant branch (RGB). For systems with longer orbital periods, this type of interaction is even more unlikely.

On the other hand, tidal stellar capture \citep{Heggie2003} could, in principle, produce binaries with different component masses and ages. However, such events are rare and occur mostly in clusters. \cite{Ivanova2005} estimated that tidal-capture binaries form at a rate of only about 1$\%$ over the lifetime of a globular cluster core. Classical Cepheids are not expected in globular clusters, and only one BIND Cepheid is close to an open cluster, so a tidal-capture origin appears highly improbable. Moreover, within a cluster, the ages, masses, and periods of Cepheids should be similar, as, for example, in NGC 1866 and NGC 2031 \citep{Pilecki2024a}.

Another possible explanation is a stellar merger \citep[see, e.g.,][]{Schneider2025}. In this scenario, the system would have originated as a triple. In such a system, the Kozai-Lidov mechanism \citep[][and references therein]{Naoz2016}, tidal forces, and standard stellar evolution may eventually lead to a merger of the inner binary, producing a rejuvenated star. The tertiary evolves separately and remains largely unaffected, serving as the true age marker of the system. The evolution before and after the merger event has to be balanced so that its product crosses the IS at the same time as the third star crosses it during its standard single-star evolution. As noted in \cite{Espinoza-Arancibia2025}, mergers are expected to introduce significant apparent age differences between components when single-star evolution is considered for both components. The same was also noted previously by \citet{Bond2018} for Polaris, and by \citet{Neilson2015} for OGLE-LMC-CEP-1812.

For systems with an age difference and mass ratios slightly below unity ($0.9<q_{\rm p}<0.95$), the interpretation is elusive. Technically, the more massive component can still be a product of binary interaction, but matching the ages becomes increasingly difficult as the masses of the inner binary components decrease relative to the previous tertiary component.
Quantitatively,  pulsation-driven mass loss may offer a viable explanation. \citet{Neilson2008} and \citet{Neilson2011} showed that Cepheids can lose 5–10$\%$ of their total mass through pulsation-driven winds. However, in the case of binary Cepheids, this phenomenon affects both components, and its final effect is differential. Moreover, it primarily affects the initially more massive and evolved component, inverting the mass ratio rather than decreasing it. The best example is the current configuration of LMC-CEP-1718, described earlier.

To distinguish among these formation scenarios, empirical constraints are crucial. In particular, an accurate mass ratio is fundamental, as each scenario predicts significantly different values of $q_{\rm p}$. Sometimes, the period change rate (PCR) is used as an evolutionary diagnostic: a positive value indicates either the first or third crossing across the IS, and a negative value indicates the second crossing \citep[see, e.g.,][]{Espinoza2022}. Moreover, the period change on the 1C should be much faster than on the third crossing. However, non-evolutionary or erratic variations can obscure the evolutionary signal \citep[][]{Rathour2025}, and relying solely on PCR measurements would require very long time baselines \citep[see, e.g.,][where a time baseline of over 100 years was used to derive PCR of a sample of LMC Cepheids]{Rodriguez2022}. Therefore, we do not consider PCR a decisive indicator. Nevertheless, its value may support one of the solutions, especially if it is already favored for another reason. Another weak indicator is the mass ratio of the individual solution. We may expect configurations with mass ratios close to unity to be more probable, as they do not require a past binary interaction. 

For two of the systems (GD-CEP-291 and SMC-CEP-1526), we could determine spectroscopic mass ratios with sufficiently low uncertainties to uniquely identify one of the available configurations. These mass ratios come from orbital solutions based on a greatly extended observational baseline and pulsational radial velocity curve coverage, compared with those shown in Paper II\footnote{In Paper II, the focus was on detecting binary motion rather than deriving accurate orbital parameters. Updated orbital solutions will be presented in a forthcoming paper in the series.}, thereby significantly reducing the degeneracy in disentangling orbital and pulsational motion. For the remaining seven systems, we default to the preferred configuration of two BL Cepheids. This is because BL evolution is up to 100 times slower than 1C evolution \citep{Smolec2026}, and the system is much more likely to be in the former configuration. However, we also discuss alternative solutions allowed by our modeling results when their likelihoods are comparable. 

BLG-CEP-067 contains a secondary component that exhibits a negative PCR at 11-sigma significance \citep{Pilecki2024a}, indicating evolution along the second IS crossing. Although we consider PCRs as a secondary diagnostic, this favors the solution in which both stars are BL Cepheids (BL+BL solution). However, it requires at least one component to be a binary interaction product, whereas the alternative solution with a BL+1C configuration does not require this, yielding an expected mass ratio close to unity. As the spectroscopic mass ratio for this system is not yet well determined, we default to the BL+BL solution; however, based on our models, there is a significant probability that the secondary is a 1C Cepheid.

GD-CEP-291 allowed 4 different configurations of 1C and BL Cepheids. The predicted mass ratios differ significantly, except for the BL+BL and the 1C+1C configurations, which yield $q_{\rm p} \sim 1$. The well-constrained spectroscopic mass ratio for this system, $q_{\rm s} = 0.84 \pm 0.04$, closely matches the 1C+BL Cepheid solution. Such a value also requires the primary to be a product of binary interaction. As the alternative solutions are at least 3$\sigma$ away from $q_{\rm s}$, we adopted the 1C+BL configuration as the most likely one. At the moment, this is the only system for which a solution from Appendix~\ref{tab:results_1st} involving a 1C Cepheid is adopted over the BL+BL configuration.

LMC-CEP-0571 yielded only a single viable solution in which both components have similar masses and ages and are evolving along the BL. Naturally, this is the most likely explanation for this system configuration. The negative PCR for the primary is very strong, which is consistent with the predicted second crossing. For the secondary, it is positive although less significant, suggesting a third crossing, which is a likely solution (see Appendix C) given the 1O pulsation mode and the possibility for $q_p>1$ within 1$\sigma$.

LMC-CEP-0835 also showed a single solution consisting of two BL Cepheids, yielding, however, a lower mass ratio of $0.88 \pm 0.02$ and an age difference between components of 50 Myr. With the long orbital period of this system ($\sim$6400 days), this means the primary component is a strong candidate for another merger-origin Cepheid. In this case, different masses are also necessary to explain the significantly different pulsation periods of the two Cepheid components. Strong negative PCR for the secondary could suggest it is on a second crossing, but the change is superposed on the light-travel time effect due to orbital motion with a long period, decreasing the significance of this conclusion.

SMC-CEP-1526 admitted three viable configurations, two of which involve 1C Cepheids. We recently improved the empirical mass-ratio measurement, obtaining $q_{\rm s} = 1.00 \pm 0.05$. This value rules out the 1C solutions and strongly favors the configuration of two BL Cepheids not affected by binary interactions.

SMC-CEP-2699 had two viable solutions, yielding configurations similar to those obtained for BLG-CEP-067. For this system, we also default to the BL+BL solution, although our models require a binary evolution origin of the primary (predicted $q_{\rm p} = 0.80 \pm 0.06$). Similarly, to LMC-CEP-0835, this $q_{\rm p}$ explains the significantly different pulsation periods of the components. The alternative is a configuration of BL+1C Cepheids of similar masses. In the latter solution, the 1C Cepheid is, strangely, predicted to be slightly more massive. However, the observed strong period decrease in the secondary is not expected for a first crosser, supporting the BL+BL solution.

SMC-CEP-2893, SMC-CEP-3115, and SMC-CEP-3674 each have three solutions for different 1C and BL Cepheid combinations. In each case, these three solutions yield significantly distinct predicted mass ratios. Given the lack of reliable spectroscopic mass ratios for these systems, we adopted the BL+BL Cepheid solutions with predicted $q_{\rm p} \sim 1$ as the most probable.

For seven systems with no reliable spectroscopic mass ratios, future measurements of these values will provide strong constraints on their evolutionary configurations, either confirming our most probable solutions or revealing even more interesting configurations involving 1C Cepheids.

\subsection{Period--mass--radius relation}\label{subsec:PMR}

\begin{figure*}
    \centering
    \includegraphics[width=\linewidth]{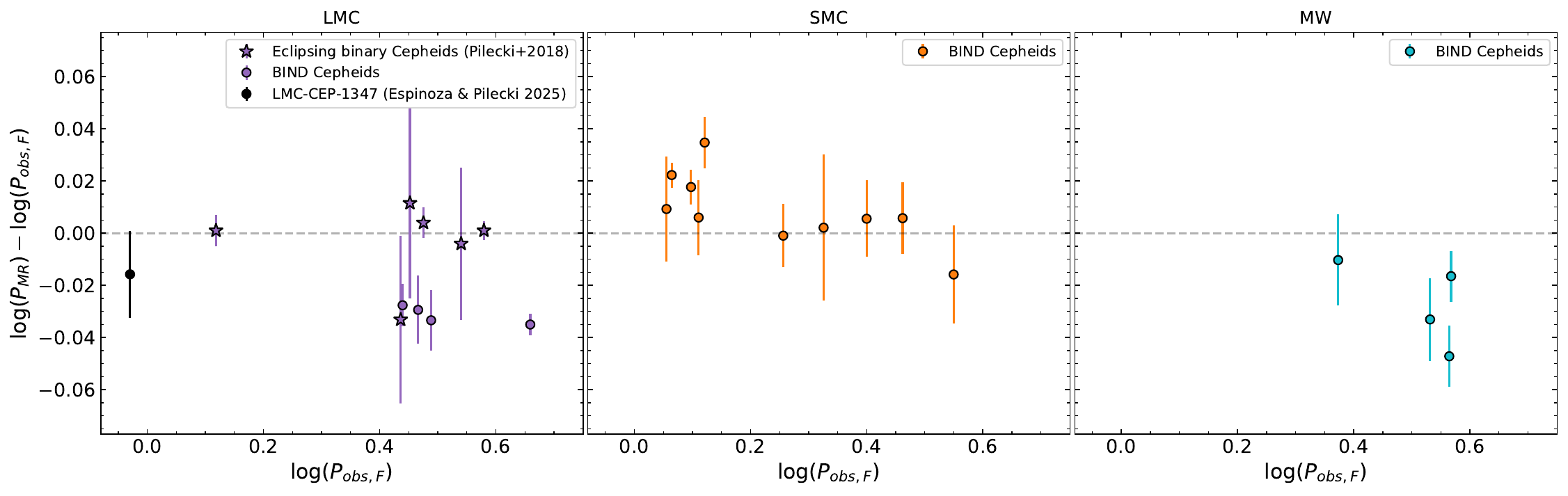}
    \caption{Differences between the period from the PMR relation obtained by \citet{Pilecki2018} and the observed period for our studied BIND Cepheids in the LMC, SMC, and MW. In the left panel, eclipsing binary Cepheids from \citet{Pilecki2018} and the LMC-CEP-1347 system \citep{Espinoza-Arancibia2025} are also included.
    }
    \label{fig:PMR-Pilecki}
\end{figure*}

Based on precisely measured parameters of five eclipsing binary Cepheids and one Type II Cepheid in the LMC, \citet{Pilecki2018} derived a PMR relation\footnote{$\log P_{\rm MR} = -1.555(35) - 0.795(44)\log M + 1.703(23)\log R$, where the mass and radius are in solar units.}. Using this relation, we compute a pulsation period ($P_{\rm MR}$) for all BIND Cepheids, adopting the mass and radius obtained with the $q$-PED method. We considered the solutions that presented two BL Cepheids, except for GD-CEP-291, for which we used the 1C+BL solution. In Fig.~\ref{fig:PMR-Pilecki}, we compared our results using this PMR relation and the observed pulsation period of the corresponding Cepheid. For 1O pulsators, we fundamentalized their periods using the equations given by \citet{Pilecki2024}. In general, the differences in period are lower than 8$\%$. Our results for LMC and MW Cepheids yielded periods that were, on average, 6$\%$ shorter than those observed. On the other hand, most the SMC Cepheids present slightly longer periods than observed, by up to 7$\%$ on average. Using only the results of this work, we computed a new $q$-PED-based PMR relation, which reads as follows:
\begin{equation}\label{eq:newPMR}
    \log P^{\rm qPED}_{\rm MR} = - 1.675(39) - 0.653(82)\log M
    + 1.741(57)\log R,
\end{equation}
where the mass and radius are in solar units. Similarly to Fig.~\ref{fig:PMR-Pilecki}, we computed the difference between the periods from this new PMR relation and the observed periods for all Cepheids. This time, the differences were smaller than 3$\%$. As pulsation periods were used as input in the $q$-PED method, the small differences observed assure internal consistency on our modeling method.

\begin{figure}
    \centering
    \includegraphics[width=\linewidth]{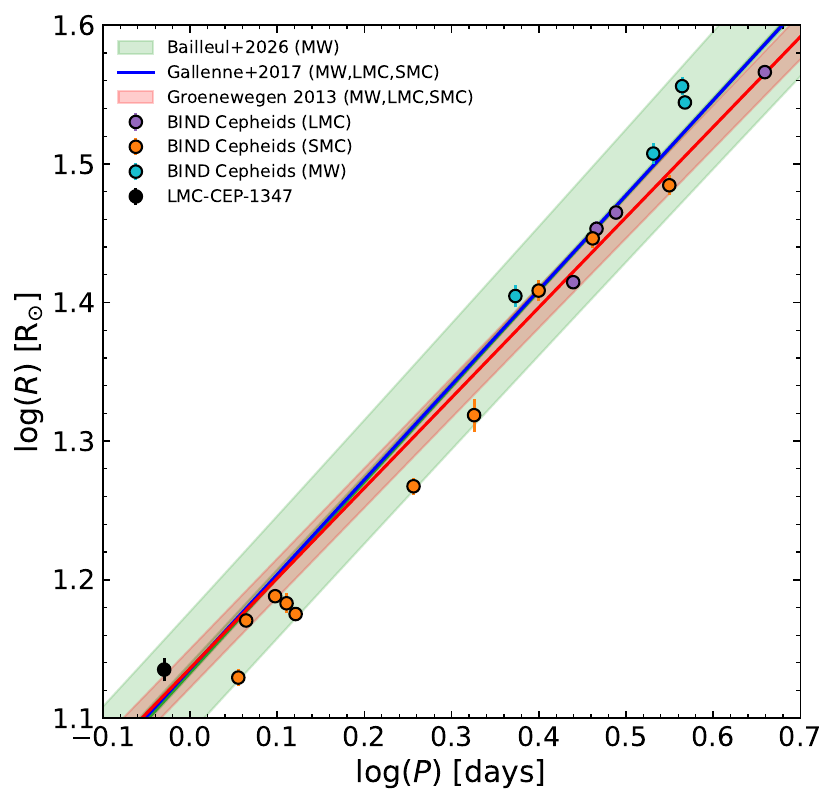}
    \caption{Radius as a function of pulsation period for our LMC, SMC, and MW BIND Cepheids. For comparison, three empirical PR relations are shown as solid lines, with their uncertainties indicated as shaded areas. The statistical uncertainties provided by \citet{Gallenne2017} are too small to be visible, while their relation practically overlaps with that of \citet{Bailleul2026}. The $q$-PED radius for LMC-CEP-1347 \citep{Espinoza-Arancibia2025} is also shown.}
    \label{fig:PR}
\end{figure}

Additionally, in Fig.~\ref{fig:PR} we compare our calculated radii and observed periods with the empirical period--radius (PR) relations from \citet{Bailleul2026}, \citet{Gallenne2017}, and \citet{Groenewegen2013}. The relation of \citet{Groenewegen2013} was computed using 1O- and F-mode Galactic Cepheids, in addition to F-mode Cepheids in the Magellanic Clouds. The \citet{Gallenne2017} relation was obtained from a sample of LMC and SMC F-mode Cepheids, and that of \citet{Bailleul2026} using only F-mode Galactic Cepheids.  Overall, our results from the modified $q$-PED method for the components of BIND Cepheids are within the uncertainties of the \citet{Bailleul2026}  relation. For the LMC sample, they are in very good agreement with the empirical PR relations, yielding differences of around 3$\%$ on average. The $q$-PED radii for our Galactic sample lie slightly, but systematically, above all relations, displaying differences of around 7$\%$ with that of \citet{Groenewegen2013} and 3$\%$ with the one from \citet{Bailleul2026}. In contrast, the majority of SMC Cepheids with periods shorter than 2.5 days show $q$-PED radii that are approximately 8\% smaller than the empirical predictions. Note, however, that the relations of \citet{Gallenne2017} and \citet{Groenewegen2013} were calibrated on Cepheids with periods longer than 2.5 days, so their applicability in this short-period regime is uncertain. Still, the $q$-PED radius of LMC-CEP-1347 \citep{Espinoza-Arancibia2025}, our only short-period LMC Cepheid, lies above all the relations.
Qualitatively, these systematic shifts in radius most probably arise from differences in metallicity between the samples. Quantitatively, however, they may also depend to some extent on the assumed physical input parameters of our evolutionary and pulsation models.

\subsection{Mass--luminosity relation}

The empirically determined parameters of Cepheids currently span only a narrow range of stellar masses and metallicities \citep{Pilecki2018}. Although our method makes use of evolution and pulsation theory, it still faces many strong empirical constraints and, as such, may provide important insights into the ML relation. In fact, masses and luminosities from the $q$-PED method are remarkably close to those from the eclipsing binary modeling, as can be seen in Table~\ref{tab:LMC-CEP-1718-comparison} and in \citet{Espinoza-Arancibia2025}. To determine the ML relation, we used the component properties from the most likely solutions: the 1C+BL solution for GD-CEP-0291 and the BL+BL solutions for the remaining systems. The fitted relation is:

\begin{equation}\label{eq:ML}
    \log M = 2.922(157)\log P + 1.285(85).
\end{equation}

\begin{figure*}
    \centering
    \sidecaption
    \includegraphics[height=0.25\textheight]{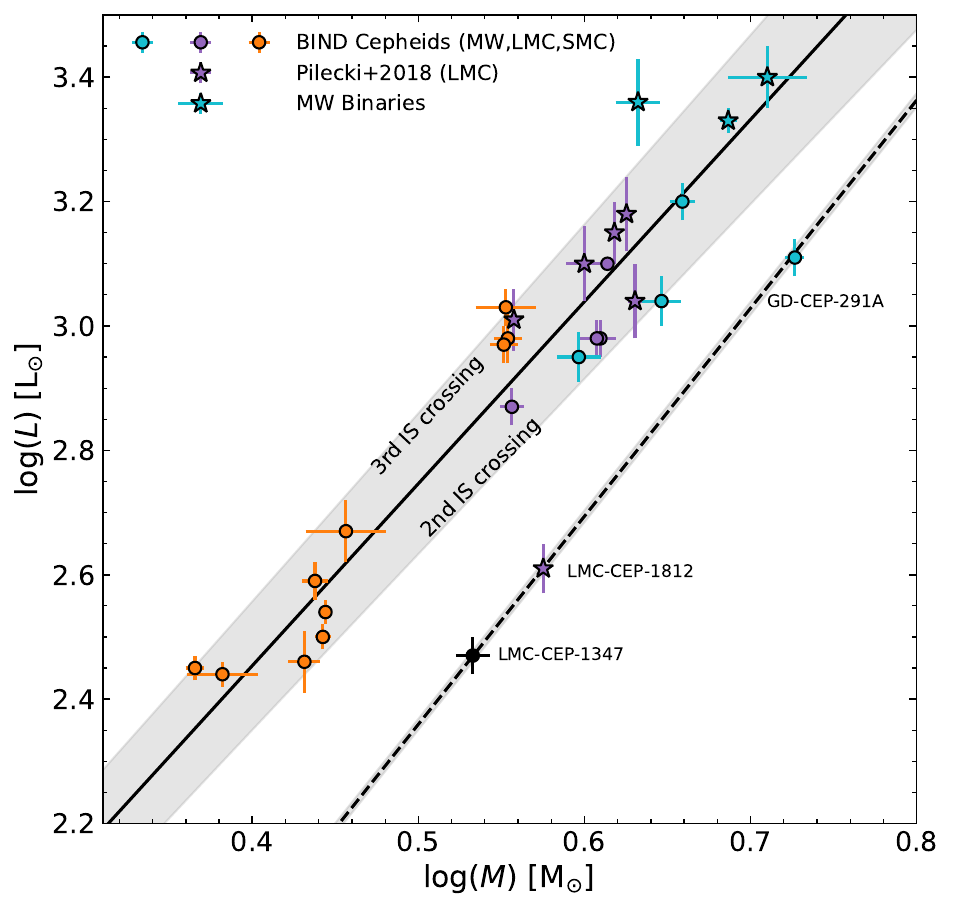}
\includegraphics[height=0.25\textheight,trim=0.9cm 0.0cm 0cm 0.0cm,
    clip]{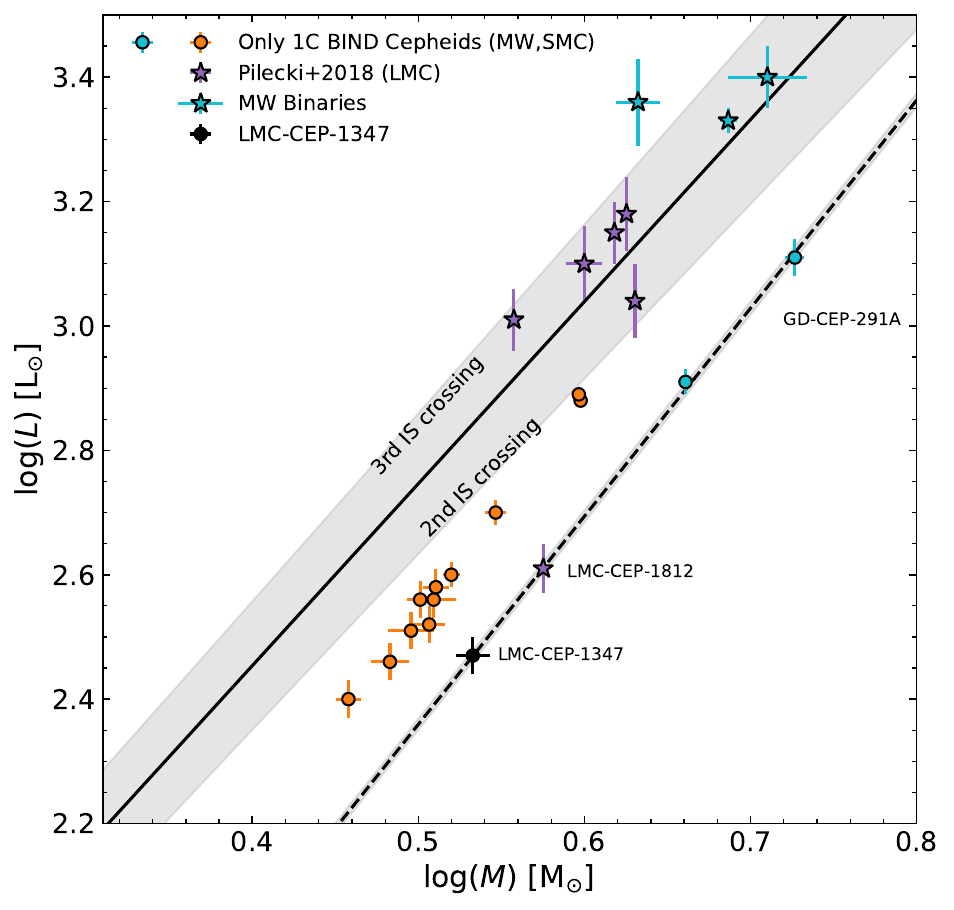}
    \caption{Luminosity as a function of mass. The solid line represents eq.~\ref{eq:ML}, while the uncertainties of this relation are shown as a shaded area. In the left panel, we present our results for solutions with two BL Cepheids, except for the GD-CEP-291 system for which the measured $q_s$ confirmed the 1C+BL solution. The right panel shows the results only for 1C Cepheids from the alternative solutions that include at least one 1C Cepheid. Stars represent empirical results for the LMC eclipsing \citep{Pilecki2018} and MW binary Cepheids \citep{Gallenne2019, Evans2024a, Gallenne2025}. Preliminary M-L relation for the first crossing Cepheids is marked with a dashed line.}
    \label{fig:ML}
\end{figure*}

In the left panel of Fig.~\ref{fig:ML}, we present our $q$-PED-based ML relation together with the results obtained in this work, as well as well-studied LMC Cepheids in eclipsing binaries \citep{Pilecki2018}, and the Galactic Cepheids: Polaris \citep{Evans2024b}, V1334 Cygni \citep{Gallenne2018}, and SU Cyg \citep{Gallenne2025}. With the exception of V1334 Cygni, all empirical measurements are consistent, within the uncertainties, with our derived ML relation.

Interestingly, we found that the determined ML relation successfully separates Cepheids in the second and third crossings of the IS. They are located below and above the solid line (eq.~\ref{eq:ML}), respectively. We confirmed this by comparing the location of the eclipsing binary Cepheids of \citet{Pilecki2018} in the ML plane with their likely evolutionary state determined by \citet{Deka2025}. The ML plane may provide a useful diagnostic of the evolutionary status when masses and luminosities are independently constrained. We summarized the inferred IS crossings of our BL solutions in Sec.~\ref{sec:conclusions}.

In the right panel of Fig.~\ref{fig:ML}, we show only the 1C Cepheids obtained in the possible configurations of our sample. These objects clearly follow a fainter ML relation than that derived for BL Cepheids. However, all SMC Cepheids still lie significantly above the relation defined by LMC-CEP-1347 \citep{Pilecki2022, Espinoza-Arancibia2025} and LMC-CEP-1812 \citep{Pilecki2018} that were identified with high confidence as 1C Cepheids. Evolutionary tracks we calculated for this range of masses, indeed show SMC stars may be more luminous by 0.1 in $\log L$, but here a shift of 0.2 is present. At the moment, we treat this inconsistency to favor the default BL+BL solutions for the SMC systems.

On the other hand, the 1C components of the MW sample appear to follow the same trend as those of LMC-CEP-1347 and LMC-CEP-1812. This is not a great surprise, given our discussion of their evolutionary configuration in Section~\ref{subsec:origins}. The spectroscopic mass ratio for GD-CEP-0291 directly indicates the 1C+BL solution, and for BLG-CEP-067, we found the configuration with a 1C Cepheid to be of comparable likelihood because it did not involve binary interactions ($q_p \sim 1$). Using the properties of the primary 1C of the GD-CEP-0291 system together with those of LMC-CEP-1347 and LMC-CEP-1812, we obtain the preliminary $q$-PED-based ML relation for the first crossing Cepheids:
\begin{equation}\label{eq:ML_1C}
    \log M = 3.299\log P + 0.712.
\end{equation}

\section{Summary and conclusions}\label{sec:conclusions}

In this work, we have extended the recently introduced $q$‐PED method to a sample of nine BIND Cepheids located in the MW, LMC, and SMC galaxies. Using dense grids of MESA evolutionary tracks and RSP linear pulsation models, combined with multi‐band photometry and distances to the systems' host galaxies, we estimated the physical parameters for both components and explored their current evolutionary status. These results significantly enlarge the sample of Cepheids with determined masses and radii, and allow us to extend the ML relation to lower masses, down to 2.3 M$_\odot$, which is much lower than typically expected for Cepheids. 

Our modeling of the benchmark eclipsing BIND Cepheid LMC‐CEP‐1718 shows that the q‐PED method can reproduce its fundamental parameters to within a few per cent, including mass, radius, luminosity, and effective temperature. However, the mass ratio remains slightly inverted relative to the empirical value, possibly indicating enhanced pulsation-driven mass loss. This validation confirms that q‐PED provides reliable physical parameters for BIND Cepheids, even in the presence of unaccounted perturbing effects.

We applied the modified $q$-PED method to the full BIND sample. Based on our models, we found that most systems admitted multiple evolutionary configurations, involving various combinations of 1C and BL Cepheids and different mass ratios. To quantify the minimum mass ratio expected for non-interacting BIND Cepheids, we constructed isochrones from our evolutionary tracks and used empirical IS boundaries from our previous studies. We found that the minimum mass ratio decreases with metallicity, reaching values as low as $q\simeq0.935$ in the SMC, and $q\simeq0.95$ in the LMC and MW. These values define the observationally constrained model-based estimates of minimum $q$ for which no significant binary interaction is necessary to explain the configuration of a BIND Cepheid system.

We then incorporated independent empirical constraints, such as spectroscopic mass ratios, to discriminate between the viable configurations. The systems LMC-CEP-0571, SMC-CEP-1526, SMC-CEP-2893, SMC-CEP-3115, and SMC-CEP-3674 are most likely composed of BL Cepheids that did not experience binary interactions. For SMC-CEP-1526, this is further supported by a spectroscopic mass ratio measurement. BLG-CEP-067, LMC-CEP-0835, and SMC-CEP-2699 are also likely systems with two BL Cepheids, but with significantly different masses, probably due to a past binary interaction. For BLG-CEP-067, however, a non-interaction solution involving a 1C Cepheid has also been considered.

For GD-CEP-0291, the spectroscopic mass ratio clearly indicates a 1C+BL solution with the primary component being a binary evolution product. Interestingly, all three confirmed first-crossing Cepheids in binary systems (apart from this, also LMC-CEP-1812 and LMC-CEP-1347) seem to be of merger origin. Although this result may be partly due to low statistics, it is hard to believe it is just a coincidence.

While a low $q_s$ values for GD-CEP-0291 directly indicates binary interaction in the system, for LMC-CEP-0835 a low mass ratio is the only solution obtained from our observationally-constrained $q$-PED modeling. This already makes 20\% of our sample affected by binary evolution processes. Assuming BL+BL solutions as the most probable ones, this value increases to 40\% with the inclusion of BLG-CEP-067 and SMC-CEP-2699, while otherwise, an increased number of rare first-crossing Cepheids would have to be accepted.

Using the derived masses and radii for the most likely solutions (by default BL+BL except the spectroscopically confirmed one), we revisited the PMR relation by comparing predicted pulsation periods with the observed ones for all BIND Cepheids in our sample. We fitted a new $q$-PED-based PMR relation to our systems, obtaining period residuals below 2$\%$ for all objects, indicating that the inclusion of lower‐mass and lower‐metallicity Cepheids improves the calibration of the PMR relation. We also compared our $q$-PED radii with empirical PR relations, finding a good agreement for the LMC systems, whereas MW Cepheids tend to be slightly larger than predicted, and SMC Cepheids with periods shorter than about 2.5 days appear systematically smaller. This discrepancy probably arises from differences in metallicity among the Cepheid host environments. However, the existing empirical calibrations are based on longer-period Cepheids, and to compare with our short-period Cepheids, they had to be extrapolated.

We fitted an ML relation to the default blue‐loop solutions, yielding a tight correlation that covers most of the available empirical masses, except for a few outliers, such as V1334 Cyg. We also provided a preliminary relation for first-crossing Cepheids based on one confirmed 1C+BL solution for BIND Cepheids, extended with our previous solutions for other 1C Cepheids. This relation lies significantly below the relation for BL Cepheids.
Unconfirmed alternative 1C solutions for the SMC BIND Cepheids lie between these two relations. This may serve as another argument against considering these solutions valid, but only spectroscopic mass ratios will give us a definitive answer.

Taken together, our results show that double Cepheids in binary systems, when analyzed with the q‐PED method and complemented by empirical constraints, offer a powerful new window on Cepheid structure and evolution. Together with spectroscopic mass ratios, these solutions provide a direct way to identify likely merger-origin and first-crossing Cepheids, which are both rare species that are otherwise hard to confirm and lack focused studies. However, $q_{\rm s}$ is well-determined only for two systems so far. PCR is another diagnostic tool for the identification of 1C Cepheids, but erratic period changes in Cepheids result in the need for even century-long baselines to unambiguously determine the crossing number.

We are now conducting extensive spectroscopic campaigns to obtain precise mass ratios and radial velocity curves for all BIND systems, including the most challenging long-period binaries. In future work, we will incorporate these new results to confidently select the most likely configuration and origin for all systems in the sample presented in this work. 

\begin{acknowledgements}
We thank the anonymous referee for the comments and suggestions. We acknowledge support from the Polish National Science Center grant SONATA BIS 2020/38/E/ST9/00486. Support for this work is provided by ANID's FONDECYT Regular grant \#1231637 and ANID's Basal project FB210003. This research has made use of NASA's Astrophysics Data System Service. 
\end{acknowledgements}

\bibliographystyle{bibtex/aa.bst}
\bibliography{main.bib}

\begin{appendix}
\onecolumn
\begin{landscape}
\section{Multi-band photometry of BIND Cepheids}\label{table:A1-photometry}
\begin{table*}[h!]
\centering 
\begin{tabular}{cccccccccc}
\hline\hline
OGLE ID & $\langle K_{\rm s} \rangle$ [mag] & $\langle H \rangle$ [mag] & $\langle J \rangle $ [mag] & $\langle I \rangle$ [mag] & $\langle RP \rangle$ [mag] & $\langle V \rangle$ [mag] & $\langle BP \rangle$ [mag] & $E_{B-V}$ [mag]$^{\star\star}$ & References \\
\hline
   BLG-CEP-067 & 12.415(10) & 12.655(10) & 13.133(10) & 14.513(20) & - & 16.328(20) & - & 0.97(2) & 1,2\\ 
   GD-CEP-0291 & 10.249(21)$^{\star}$ & 10.515(23)$^{\star}$ & 11.186(23)$^{\star}$ & 12.700(20) & - & 14.654(20) & - & 1.06(2) & 1,3\\
   LMC-CEP-0571 & 13.785(28) & 13.842(29) & 14.224(35) & 14.878(20) & - & 15.658(20) & - & 0.19(2) & 4,5\\
   LMC-CEP-0835 & 13.643(85) & 13.647(35) & 13.937(44) & 14.491(20) & - & 15.238(20) & - & 0.04(1) & 4,5\\
   LMC-CEP-1718 & 13.564(5) & 13.614(46) & 13.944(17) & 14.519(20) & - & 15.202(20) & - & 0.13(1) & 4,6,7\\
   SMC-CEP-1526 & 15.406(9) & 15.414(17) & 15.783(4) & - & 16.183(20) & - & 16.849(20) & 0.05(2) & 4,8,9\\
   SMC-CEP-2699 & 14.519(5) & 14.531(10) & 14.820(3) & 15.403(20) & - & 16.100(20) & - & 0.07(2) & 4,8,9\\
   SMC-CEP-2893 & 15.678(8) & 15.720(15) & 15.953(6) & - & 16.361(20) & - & 16.928(20) & 1.04(6) & 4,8,9\\
   SMC-CEP-3115 & 15.530(14) & 15.561(15) & 15.823(5) & - & 16.190(20) & - & 16.690(20) & 0.02(1) & 4,8,9\\
   SMC-CEP-3674 & 14.351(3) & 14.392(10) & 14.660(3) &15.128(20) & - & 15.772(20) & - & 0.04(1) & 4,8,9\\
\hline
\end{tabular}
\tablefoot{$^{\star}$: Mean magnitudes were not available; the photometry used was acquired at a pulsation phase of 0.4, thus close to the mean value. To take into account this difference, we considered an additional error of 0.05 mag.\\
$^{\star\star}$: Reddening obtained from the multiband method for the BL+BL solution. The values, however, do not differ by more than 0.01 mag among different system configurations.\\
Magnitudes in the Visible and Infrared Survey Telescope for Astronomy (VISTA) system were converted to the Two Micron All-Sky Survey (2MASS) system using the equations of \citet{Gonzales2018}. The $H$-band magnitude in the infrared survey facility (IRSF) system was converted to the 2MASS system using the equations provided in \citet{Kato2007}.\\
For three systems, instead of OGLE magnitudes, we used Gaia RP and BP bands \citep{Gaia2023}.\\
\textbf{References} (1) \citet{Udalski2018}, (2) \citet{Herpich2021}, (3) \citet{Cutri2003}, (4) \citet{Soszynski2017}, (5) \citet{Breuval2021}, (6) \citet{Ripepi2022}, (7) \citet{Macri2015}, (8) \citet{Ripepi2016}, (9) \citet{Ita2018}}
\end{table*}
\end{landscape}

\newpage

\begin{landscape}
\section{Averaged physical parameters of BIND Cepheids.}\label{tab:results}
\begin{longtable}{ccccccccccc}
\hline\hline
OGLE-ID &Component& Mode & $Z$ & $M$ [M$_\odot$] & $T_{\rm eff}$ [K]& $\log L$ [L$_\odot$] & $\log g$ [cm/s$^2$]&Age [Myr]&$R$ [R$_\odot$]&$q$\\
\hline

\multirow{2}{*}{BLG-CEP-067}& A & 1O & 0.006/7/8 & 4.56(8) & 6170(88) & 3.20(3) & 2.02(1) & 120(10) & 35.02(36) & 0.87(3)\\
& B & 1O & 0.006/7/8 & 3.95(12) & 6269(113) & 2.95(4) & 2.23(1) & 170(10) & 25.39(48) & 0.87(3)\\\hline

\multirow{2}{*}{GD-CEP-291} & A & F & 0.009/10/11 & 4.61(10) & 5946(107) & 3.12(3) & 2.04(1) & 130(10) & 34.46(51) & 1.00(3)\\
& B & F & 0.009/10/11 & 4.59(8) & 5980(110) & 3.09(4) & 2.07(1) & 130(10) & 33.21(45) & 1.00(3)\\\hline

\multirow{2}{*}{LMC-CEP-0571} & A & F & 0.006/7/8 & 4.07(9) & 5972(97) & 2.98(3) & 2.13(1) & 160(10) & 29.17(34) & 1.00(2)\\
& B & 1O & 0.006/7/8 & 4.05(10) & 6051(84) & 2.98(3) & 2.15(1) & 160(10) & 28.39(38) & 1.00(2)\\\hline

\multirow{2}{*}{LMC-CEP-0835}& A & F & 0.005/6 & 4.11(4) & 5710(30) & 3.10(1) & 1.93(1) & 160(10) & 36.83(12) & 0.88(2)\\
& B & F & 0.005/6 & 3.60(6) & 5939(71) & 2.87(3) & 2.18(1) & 210(10) & 25.98(20) & 0.88(2)\\\hline

\multirow{2}{*}{SMC-CEP-1526} & A & F & 0.001/2 & 2.74(5) & 6006(79) & 2.59(3) & 2.35(1) & 380(20) & 18.51(26) & 0.99(3)\\ 
& B & F & 0.001/2 & 2.70(06) & 6132(141) & 2.46(5) & 2.51(1) & 370(20) & 15.24(25) & 0.99(3)\\\hline

\multirow{2}{*}{SMC-CEP-2699} & A & 1O & 0.001/2/3 & 3.57(15) & 6020(79) & 3.03(3) & 2.03(1) & 200(20) & 30.52(50) & 0.80(6)\\ 
& B & F & 0.001/2/3 & 2.86(16) & 5927(136) & 2.67(5) & 2.27(2) & 340(50) & 20.83(57) & 0.80(6)\\\hline

\multirow{2}{*}{SMC-CEP-2893} & A & F & 0.001 & 2.32(3) & 6157(48) & 2.45(2) & 2.46(1) & 550(20) & 14.97(18) & 1.04(6)\\
& B & F & 0.001 & 2.41(12) & 6424(46) & 2.44(2) & 2.57(1) & 490(70) & 13.47(19) & 1.04(6)\\\hline

\multirow{2}{*}{SMC-CEP-3115} & A & F & 0.002 & 2.78(2) & 6365(65) & 2.54(2) & 2.51(1) & 350(10) & 15.42(13) & 1.00(1)\\ 
& B & F & 0.002 & 2.77(3) & 6362(72) & 2.50(2) & 2.55(1) & 350(10) & 14.81(6) & 1.00(1)\\\hline

\multirow{2}{*}{SMC-CEP-3674} & A & F & 0.001/2/3 & 3.58(7) & 6099(117) & 2.98(4) & 2.11(1) & 200(10) & 27.94(45) & 1.00(3)\\
& B & 1O & 0.001/2/3 & 3.56(7) & 6324(82) & 2.97(3) & 2.18(1) & 200(10) & 25.62(45) & 1.00(3)\\\hline

\end{longtable}
\tablefoot{Averaged physical parameters of BIND Cepheids obtained with the $q$-PED method, considering only the solutions where both components are BL Cepheids.}
\end{landscape}

\newpage

\begin{landscape}
\section{Additional solutions obtained with the $q$-PED method}\label{tab:results_1st}
\begin{longtable}{cccccccccccc}
\hline\hline
OGLE-ID &Component& Mode & $Z$ & $M$ [M$_\odot$] & $T_{\rm eff}$ [K]& $\log L$ [L$_\odot$] & $\log g$ [cm/s$^2$]&Age [Myr]&$R$ [R$_\odot$]&$q$&Evol. phase\\
\hline

\multirow{2}{*}{BLG-CEP-067} & A & 1O & 0.007/8 & $4.70(6)$  & $6194(109)$ & $3.19(3)$ & $2.04(1)$ & $110(10)$ & $34.64(37)$ & $0.97(2)$ & BL \\
                              & B & 1O & 0.007/8 & $4.58(5)$  & $6057(57)$  & $2.91(2)$ & $2.27(1)$ & $110(10)$ & $26.18(40)$ & $0.97(2)$ & 1C \\
\hline
\multirow{2}{*}{GD-CEP-291}   & A & F  & 0.009/10/11 & $4.54(9)$  & $5900(115)$ & $3.09(4)$ & $2.04(1)$ & $130(10)$ & $33.90(52)$ & $1.15(3)$ & BL \\
                              & B & F  & 0.009/10/11 & $5.24(9)$  & $5823(94)$ & $3.07(3)$ & $2.10(1)$ & $90(10)$  & $34.26(50)$ & $1.15(3)$ & 1C \\
\hline
\multirow{2}{*}{GD-CEP-291}   & A & F  & 0.009/10/11 & $5.23(5)$  & $5734(45)$ & $3.07(2)$ & $2.07(1)$ & $90(10)$ & $35.28(24)$ & $0.98(2)$ & 1C \\
                              & B & F  & 0.009/10/11 & $5.10(6)$  & $5765(52)$ & $3.04(2)$ & $2.11(1)$ & $90(10)$  & $33.47(25)$ & $0.98(3)$ & 1C \\
\hline
\multirow{2}{*}{GD-CEP-291}
                              & \textbf{A} & \textbf{F}  & \textbf{0.009/10/11} & $\mathbf{5.33(7)}$ & $\mathbf{5791(85)}$  & $\mathbf{3.11(3)}$ & $\mathbf{2.06(1)}$ & $\mathbf{80(10)}$ & $\mathbf{35.98(53)}$ & $\mathbf{0.83(3)}$ & \textbf{1C} \\
                              & \textbf{B} & \textbf{F}  & \textbf{0.009/10/11} & $\mathbf{4.43(12)}$ & $\mathbf{5908(119)}$ & $\mathbf{3.04(4)}$ & $\mathbf{2.08(1)}$ & $\mathbf{140(10)}$ & $\mathbf{32.17(56)}$ & $\mathbf{0.83(3)}$ & \textbf{BL} \\
\hline
\multirow{2}{*}{SMC-CEP-1526} & A & F  & 0.001     & $3.31(4)$  & $5867(51)$  & $2.60(2)$ & $2.38(1)$ & $200(10)$ & $19.75(17)$ & $0.69(3)$ & 1C \\
                              & B & F  & 0.001     & $2.28(7)$  & $6192(107)$ & $2.41(4)$ & $2.50(1)$ & $580(40)$ & $14.15(25)$ & $0.69(3)$ & BL \\
\hline
\multirow{2}{*}{SMC-CEP-1526} & A & F  & 0.001/2   & $2.65(10)$  & $5999(88)$  & $2.58(3)$ & $2.35(1)$ & $410(40)$ & $18.27(24)$ & $1.15(4)$ & BL \\
                              & B & F  & 0.001/2   & $3.04(8)$  & $6028(73)$  & $2.46(3)$ & $2.54(1)$ & $250(10)$ & $15.72(23)$ & $1.15(4)$ & 1C \\
\hline
\multirow{2}{*}{SMC-CEP-2699} & A & 1O & 0.001/2   & $3.38(5)$  & $5987(44)$  & $3.00(2)$ & $2.03(1)$ & $220(20)$ & $29.80(24)$ & $1.04(2)$ & BL \\
                              & B & F  & 0.001/2   & $3.52(5)$  & $5812(58)$  & $2.70(2)$ & $2.30(1)$ & $170(10)$ & $22.44(30)$ & $1.04(2)$ & 1C \\
\hline
\multirow{2}{*}{SMC-CEP-2893} & A & F  & 0.001     & $3.17(6)$  & $6286(96)$  & $2.56(3)$ & $2.53(1)$ & $220(10)$ & $16.18(14)$ & $0.71(3)$ & 1C \\
                              & B & F  & 0.001     & $2.24(5)$  & $6243(137)$ & $2.35(5)$ & $2.57(1)$ & $600(50)$ & $12.96(17)$ & $0.71(3)$ & BL \\
\hline
\multirow{2}{*}{SMC-CEP-2893} & A & F  & 0.001     & $2.49(9)$  & $6396(67)$  & $2.52(2)$ & $2.49(01)$ & $460(50)$ & $14.95(23)$ & $1.16(4)$ & BL \\
                              & B & F  & 0.001     & $2.87(5)$  & $6109(85)$  & $2.40(3)$ & $2.60(01)$ & $280(10)$ & $14.25(18)$ & $1.16(4)$ & 1C \\
\hline
\multirow{2}{*}{SMC-CEP-3115} & A & F  & 0.001/2/3 & $3.23(10)$  & $6349(72)$  & $2.56(3)$ & $2.55(1)$ & $220(10)$ & $15.86(25)$ & $0.97(3)$ & 1C \\
                              & B & F  & 0.001/2/3 & $3.13(10)$  & $6341(82)$  & $2.51(3)$ & $2.59(1)$ & $240(10)$ & $14.97(26)$ & $0.97(3)$ & 1C \\
\hline
\multirow{2}{*}{SMC-CEP-3115} & A & F  & 0.001/2/3 & $3.24(6)$  & $6358(64)$  & $2.58(3)$ & $2.53(2)$ & $220(10)$ & $16.23(32)$ & $0.77(6)$ & 1C \\
                              & B & F  & 0.001/2/3 & $2.50(19)$  & $6363(85)$  & $2.46(4)$ & $2.54(1)$ & $460(90)$ & $14.11(51)$ & $0.77(6)$ & BL \\
\hline
\multirow{2}{*}{SMC-CEP-3115} & A & F  & 0.002     & $2.78(4)$  & $6371(62)$  & $2.53(2)$ & $2.52(1)$ & $350(10)$ & $15.28(25)$ & $1.16(2)$ & BL \\
                              & B & F  & 0.002     & $3.21(7)$  & $6344(72)$  & $2.52(3)$ & $2.58(1)$ & $230(10)$ & $15.24(31)$ & $1.16(2)$ & 1C \\
\hline
\multirow{2}{*}{SMC-CEP-3674} & A & F  & 0.001     & $3.96(3)$  & $5762(25)$  & $2.88(1)$ & $2.15(1)$ & $140(10)$ & $28.23(21)$ & $0.92(2)$ & 1C \\
                              & B & 1O & 0.001     & $3.64(7)$  & $6402(18)$  & $3.00(1)$ & $2.17(1)$ & $180(10)$ & $26.04(43)$ & $0.92(2)$ & BL \\
\hline
\multirow{2}{*}{SMC-CEP-3674} & A & F  & 0.001     & $3.64(15)$  & $6224(44)$  & $3.02(2)$ & $2.11(2)$ & $180(20)$ & $28.17(58)$ & $1.09(5)$ & BL \\
                              & B & 1O & 0.001     & $3.95(3)$  & $6021(39)$  & $2.89(1)$ & $2.22(1)$ & $140(10)$ & $25.91(33)$ & $1.09(5)$ & 1C \\\hline

\end{longtable}
\tablefoot{Averaged physical parameters of the additional solutions for BIND Cepheids obtained using the $q$-PED method. These solutions consider combinations of 1C and BL Cepheids. The solution for GD-CEP-291, featuring a 1C and a BL Cepheid, is highlighted in bold, as it is confirmed by the measured spectroscopic mass ratio.}
\end{landscape}
\twocolumn
\begin{figure*}
\section{CMDs showing the obtained BL solutions}
    \centering
    \includegraphics[height=0.22\textheight]{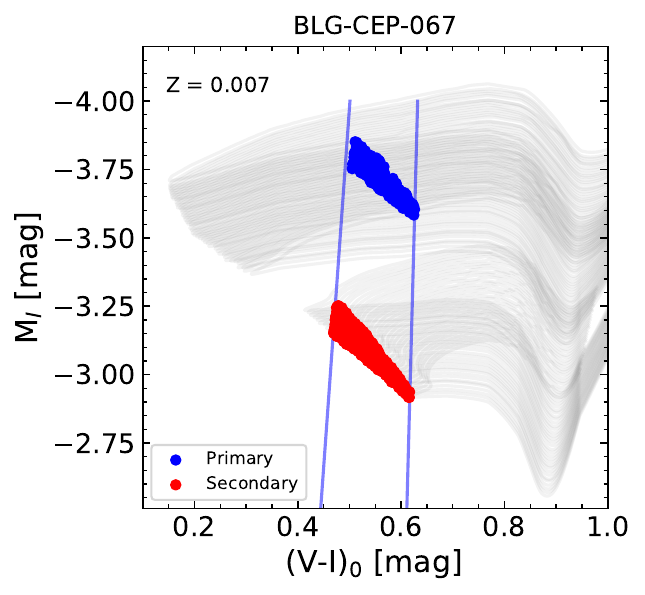}
    \includegraphics[height=0.22\textheight]{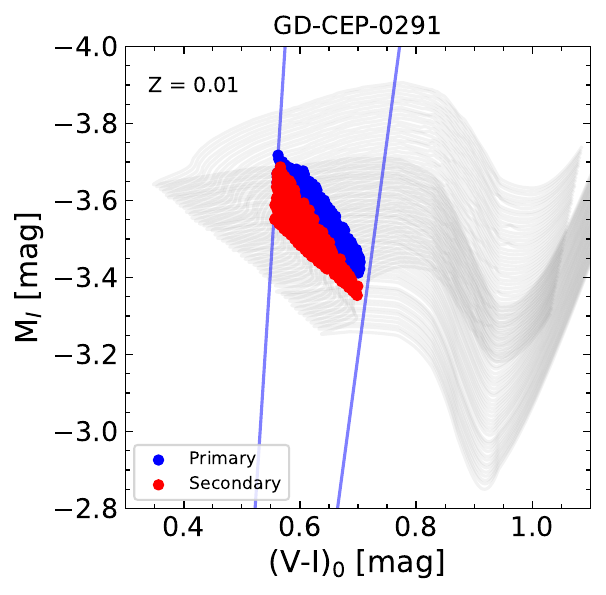}
    \includegraphics[height=0.22\textheight]{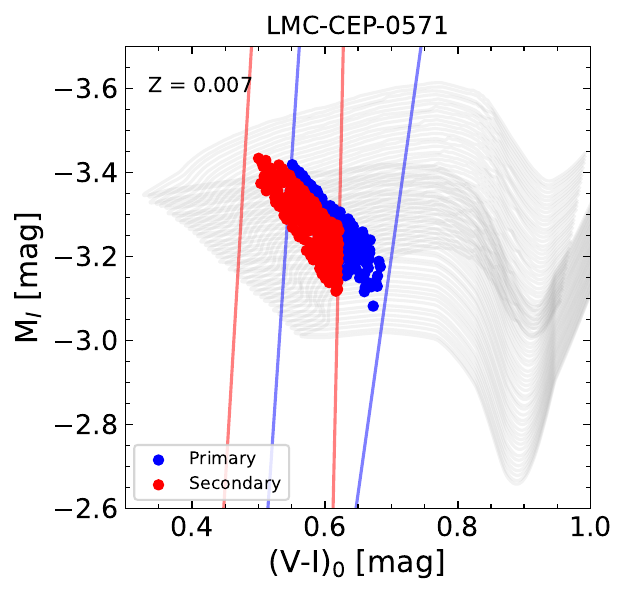}
    \includegraphics[height=0.22\textheight]{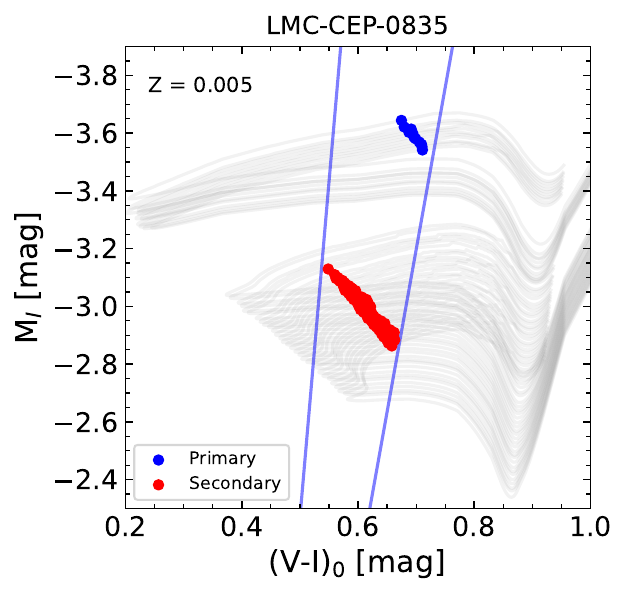}
    \includegraphics[height=0.22\textheight]{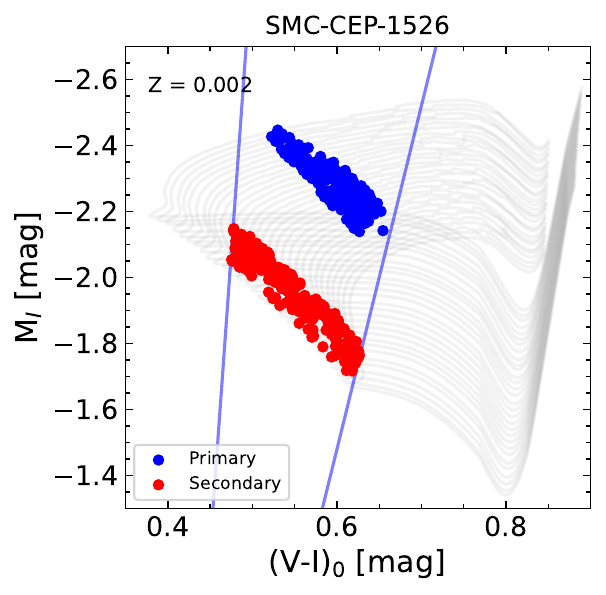}
    \includegraphics[height=0.22\textheight]{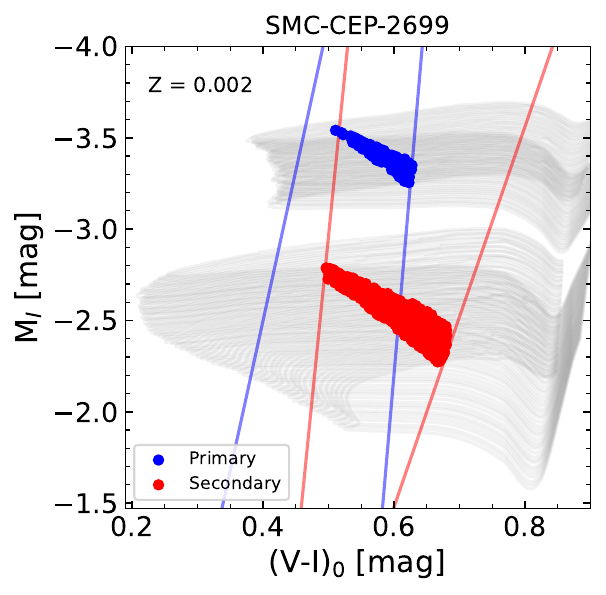}
    \includegraphics[height=0.22\textheight]{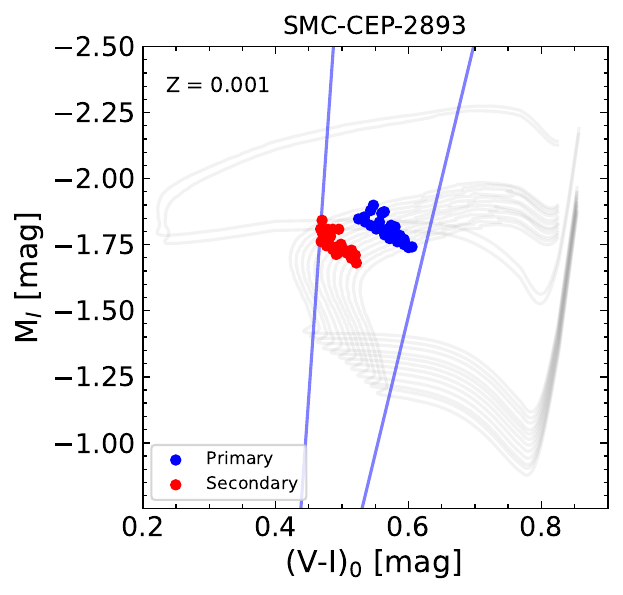}
    \includegraphics[height=0.22\textheight]{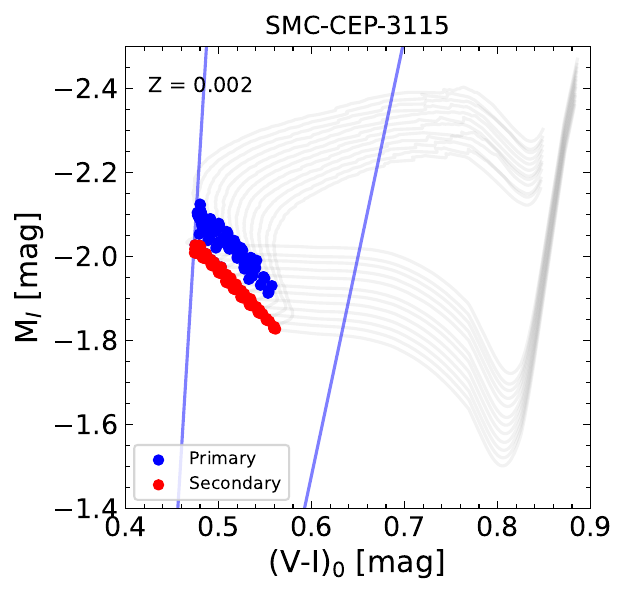}
    \includegraphics[height=0.22\textheight]{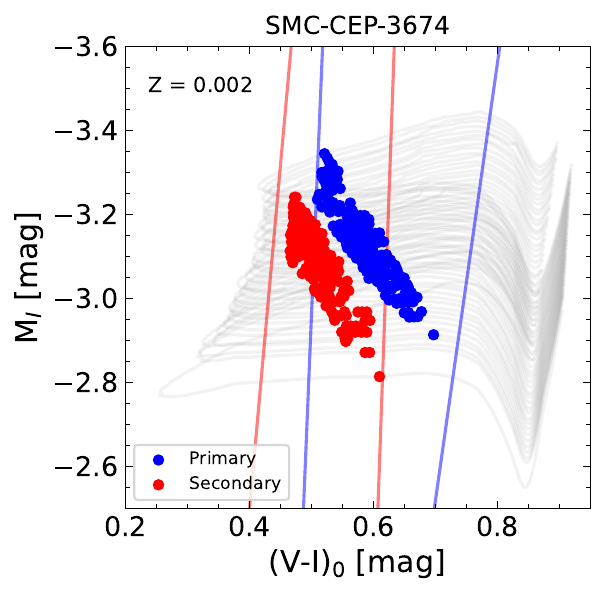}
    \caption{Color-magnitude diagrams showing all valid pulsation models for both components (blue and red points) corresponding to the BL solutions of the BIND Cepheid sample, for one value of metallicity each. The gray lines show the evolutionary tracks from which the models were derived. The blue and red lines are the 1O and F empirical instability strip edges from \citet{Espinoza-Arancibia2024}, for the primary and secondary Cepheids, depending on their respective pulsation modes.}
    \label{fig:all_BL}
\end{figure*}

\begin{figure*}
\section{CMDs showing the obtained solutions involving 1C Cepheids}
    \centering
    \includegraphics[height=0.22\textheight]{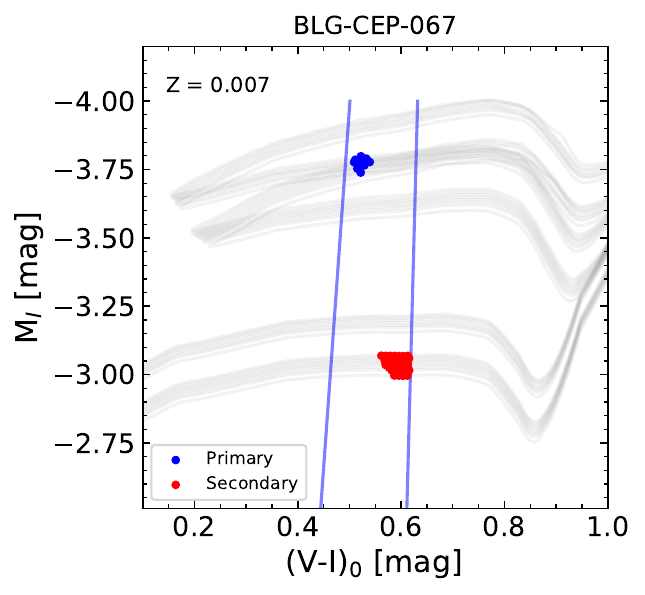}
    \includegraphics[height=0.22\textheight]{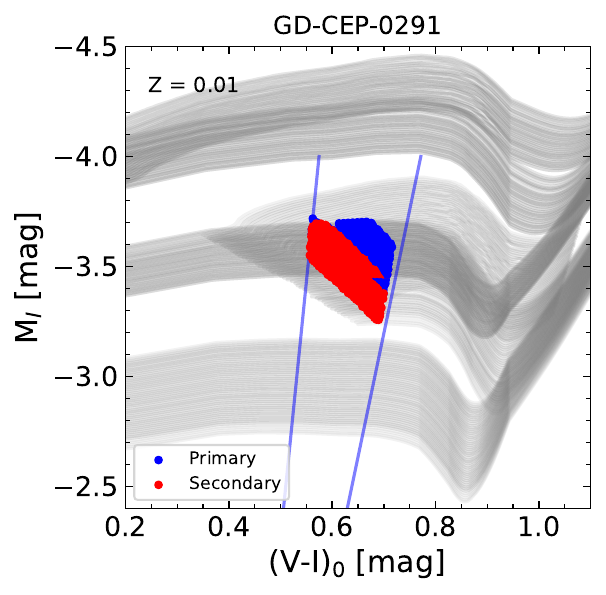}
    \includegraphics[height=0.22\textheight]{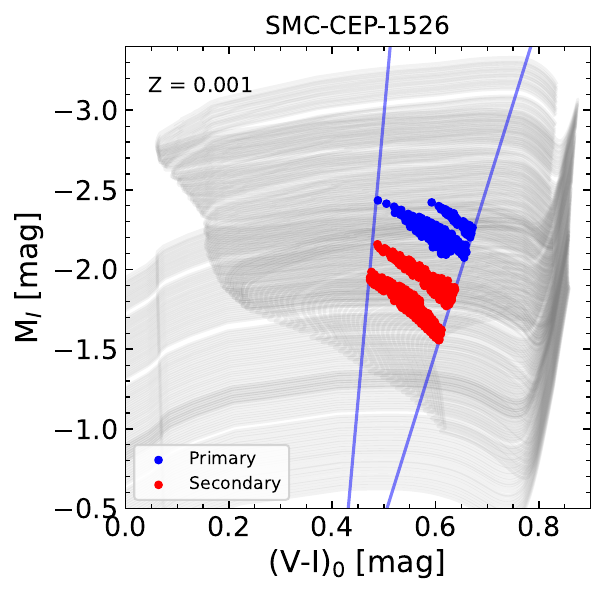}
    \includegraphics[height=0.22\textheight]{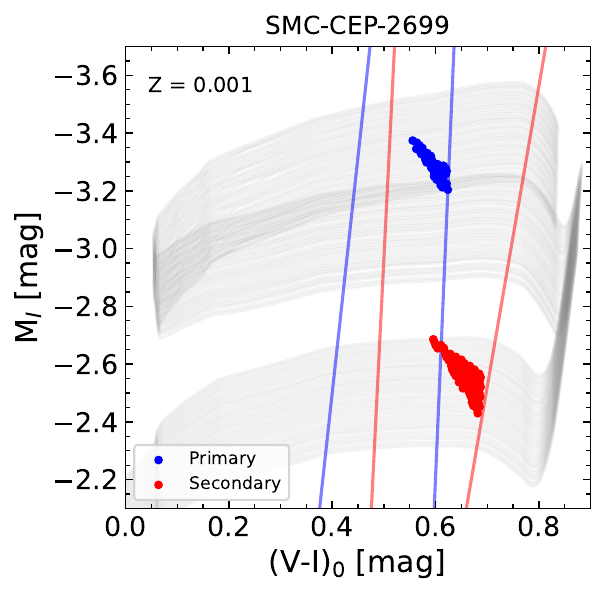}
    \includegraphics[height=0.22\textheight]{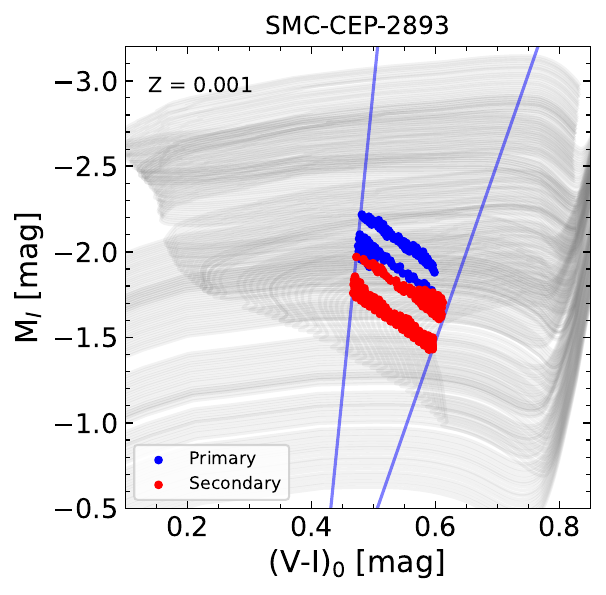}
    \includegraphics[height=0.22\textheight]{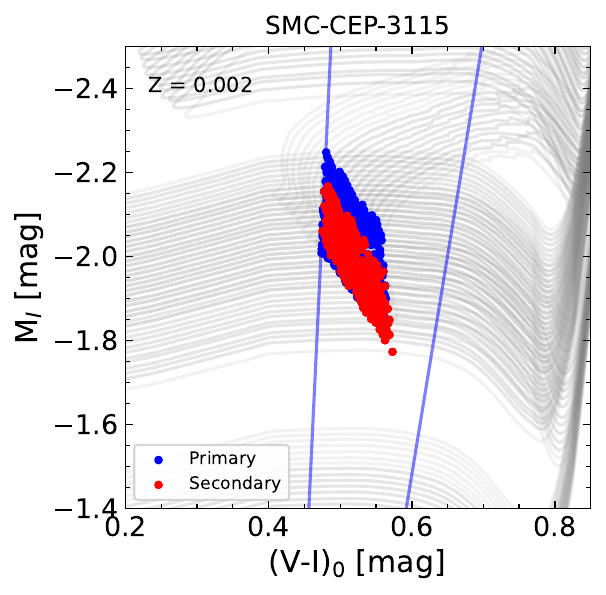}
    \includegraphics[height=0.22\textheight]{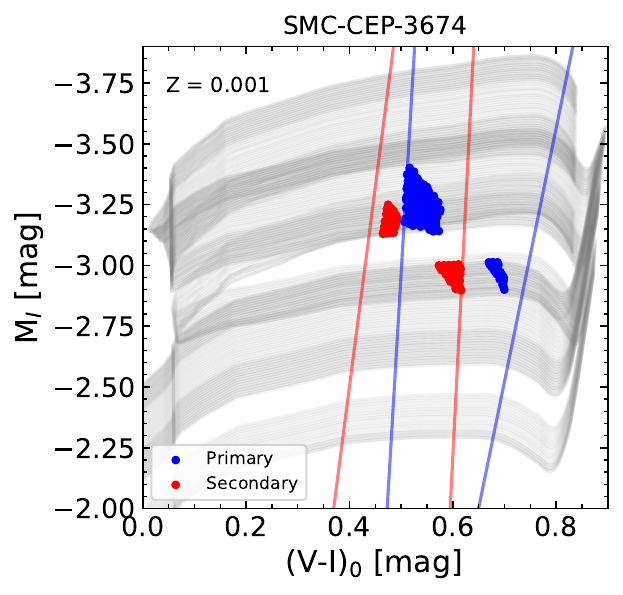}
    \caption{The same as \ref{fig:all_BL}, for the additional 1C solutions obtained for each system.}
    \label{fig:CMDs_1st}
\end{figure*}

\begin{figure*}
\section{Multi-band fit obtained for BL solutions}
    \centering
    \includegraphics[height=0.22\textheight]{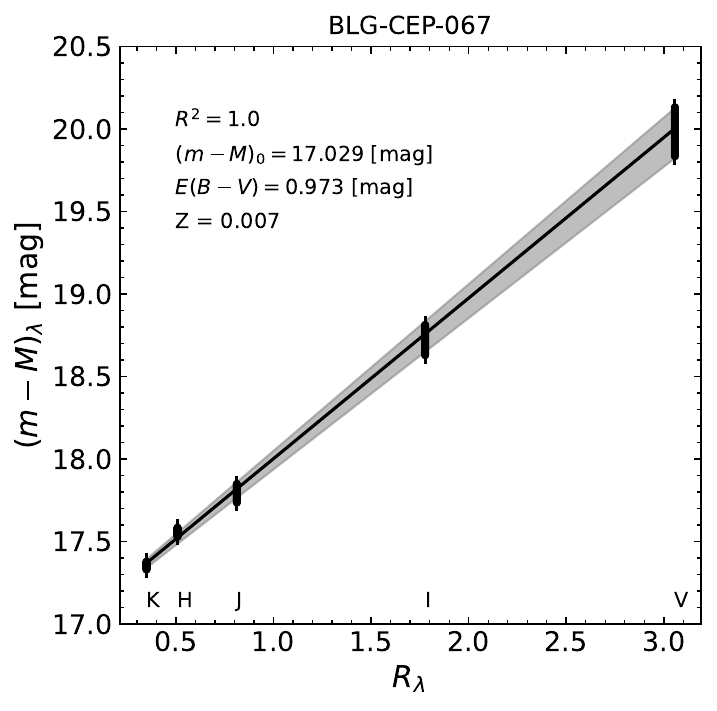}
    \includegraphics[height=0.22\textheight]{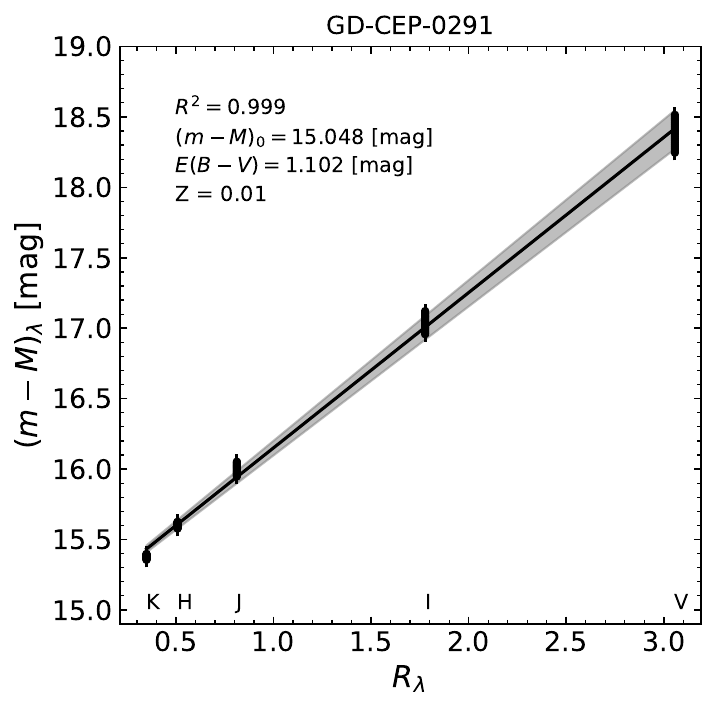}
    \includegraphics[height=0.22\textheight]{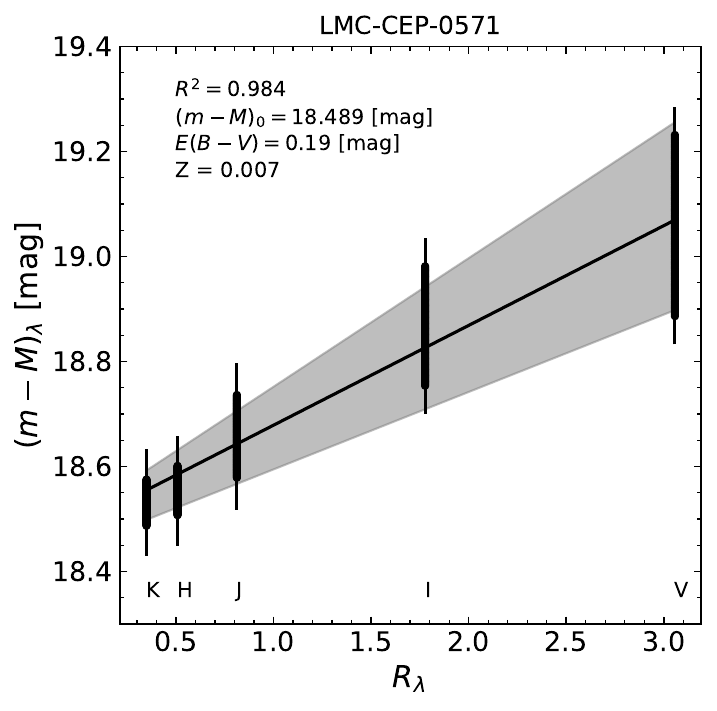}
    \includegraphics[height=0.22\textheight]{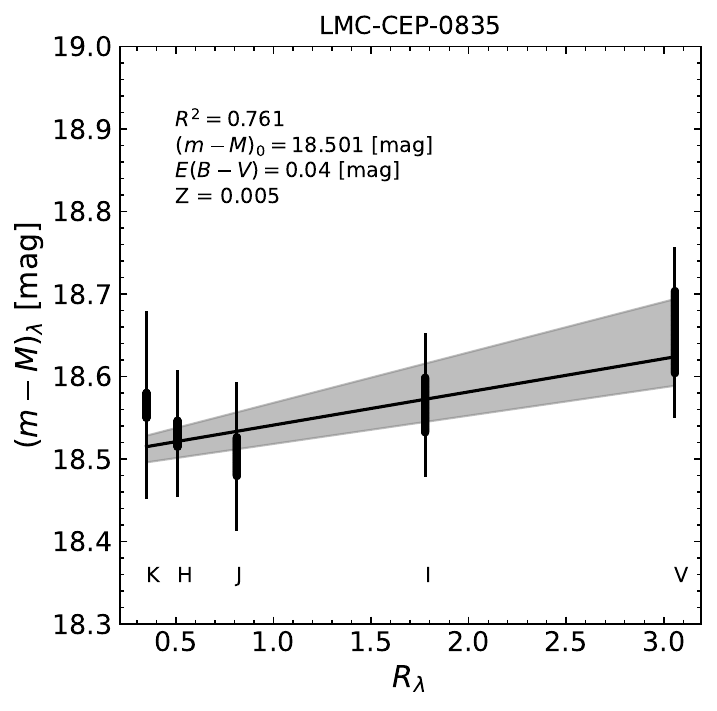}
    \includegraphics[height=0.22\textheight]{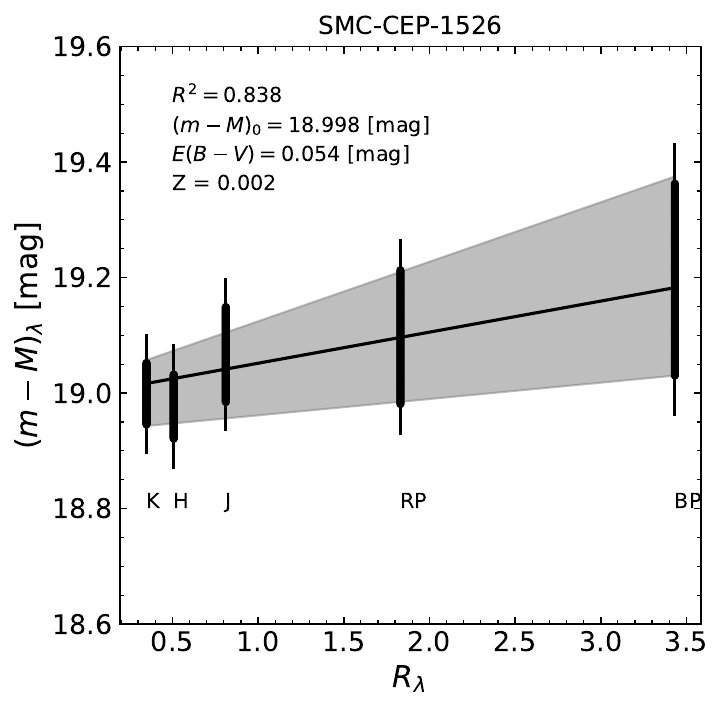}
    \includegraphics[height=0.22\textheight]{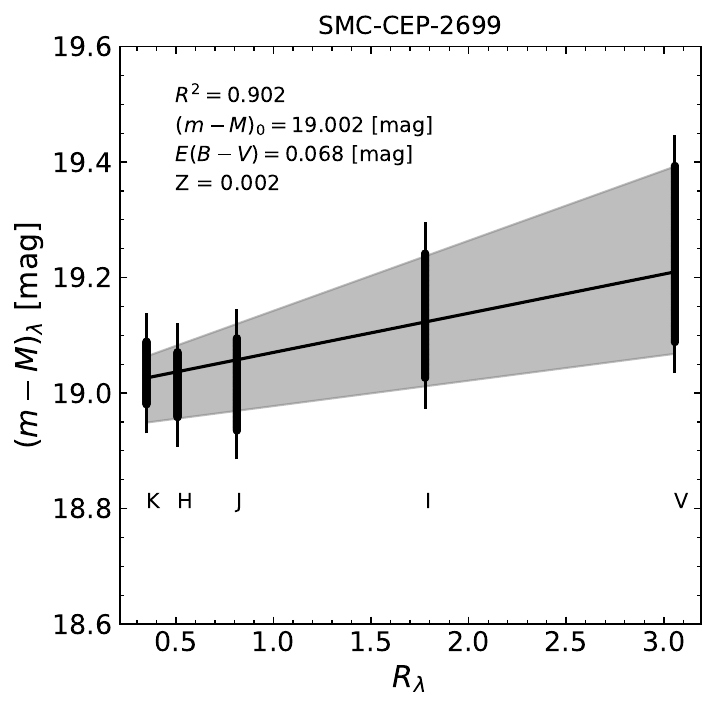}
    \includegraphics[height=0.22\textheight]{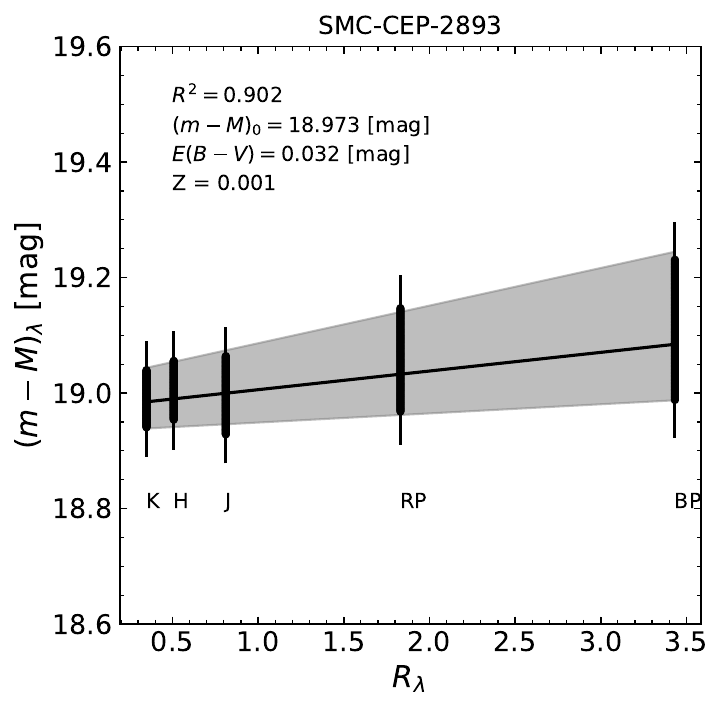}
    \includegraphics[height=0.22\textheight]{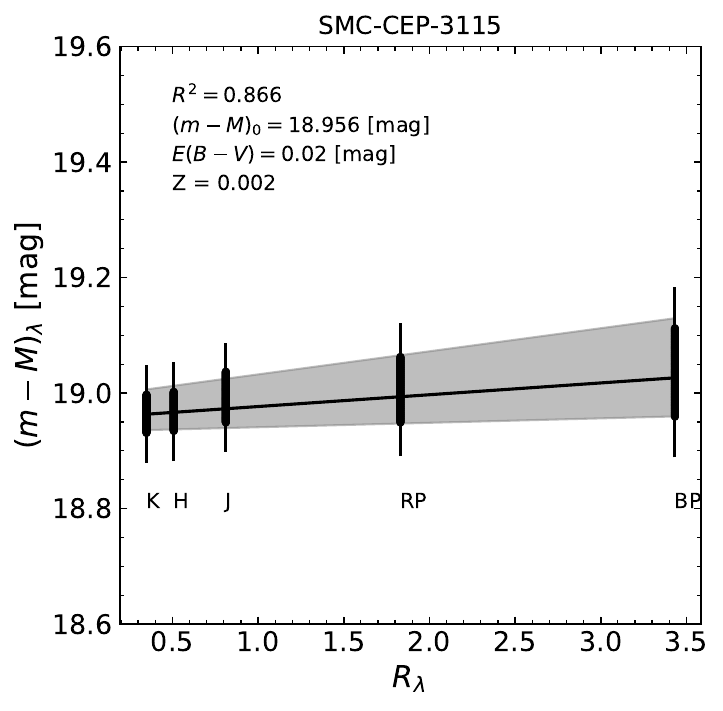}
    \includegraphics[height=0.22\textheight]{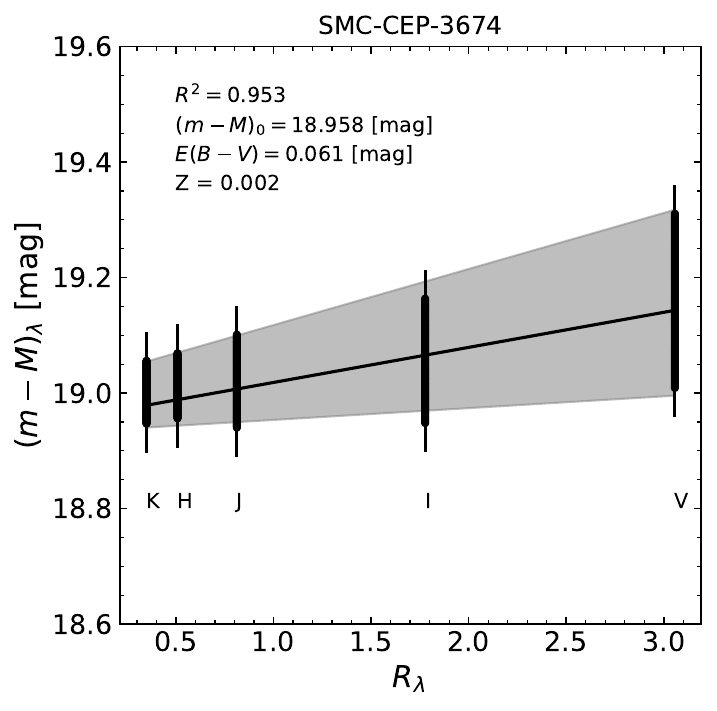}
    \caption{Multi-band fit obtained for all systems' solutions with two BL Cepheids. The black points represent the reddened distance moduli in five photometric filters, which are specified at the bottom of each panel, for all valid models shown in Fig. \ref{fig:all_BL}. The gray area indicates the range of all linear fits obtained, while the black line shows the fit with the average slope and intercept.}
    \label{fig:multiband}
\end{figure*}

\begin{figure*}
\section{HRD of MESA isochrones}
    \centering
    \includegraphics[width=\hsize]{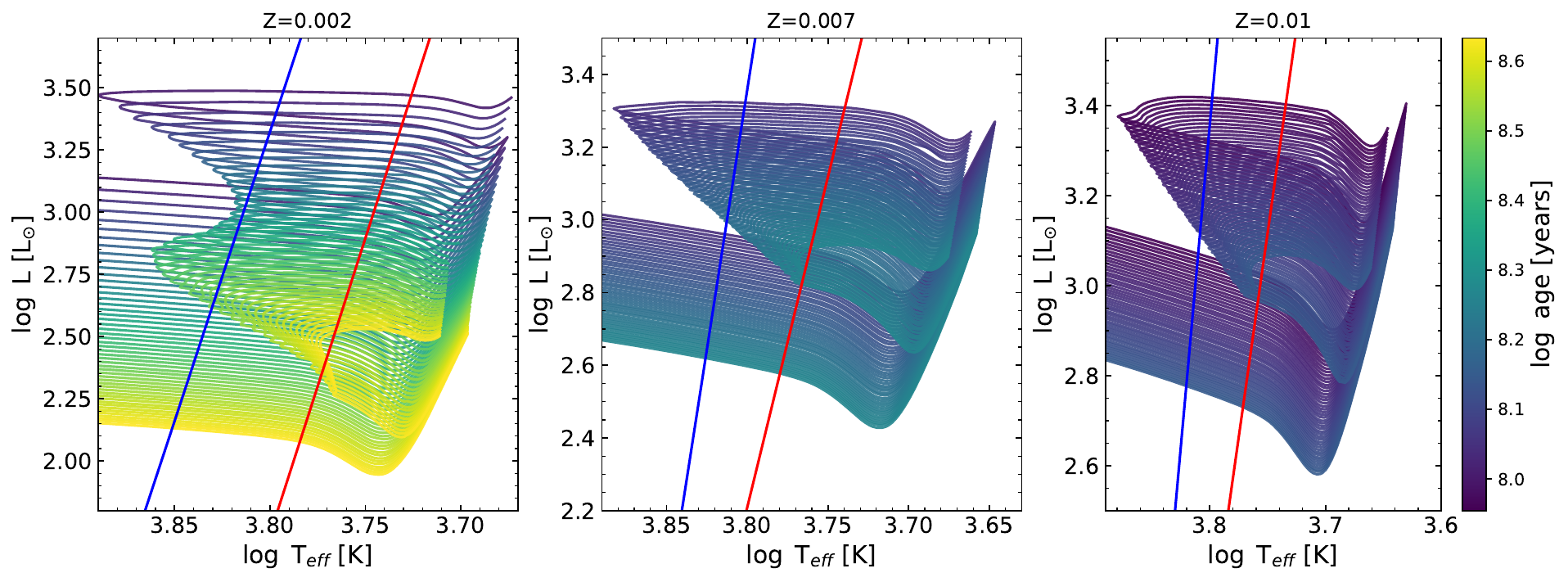}
    \caption{Hertzsprung–Russell diagram displaying grids of isochrones constructed using MESA models for three metallicities, $Z=0.002, 0.007$, and $0.01$, on the left, middle, and right panels, respectively. The colors of each isochrone represent its age. Additionally, the blue and red lines are the empirical IS edges determined in \citet{Espinoza-Arancibia2024, Espinoza2025} for the SMC (left panel) and the LMC (middle and right panels).}
    \label{fig:Iso-IS}
\end{figure*}

\begin{figure*}
\section{Differences between the new PMR relation and BIND Cepheids}
    \centering
    \includegraphics[width=\linewidth]{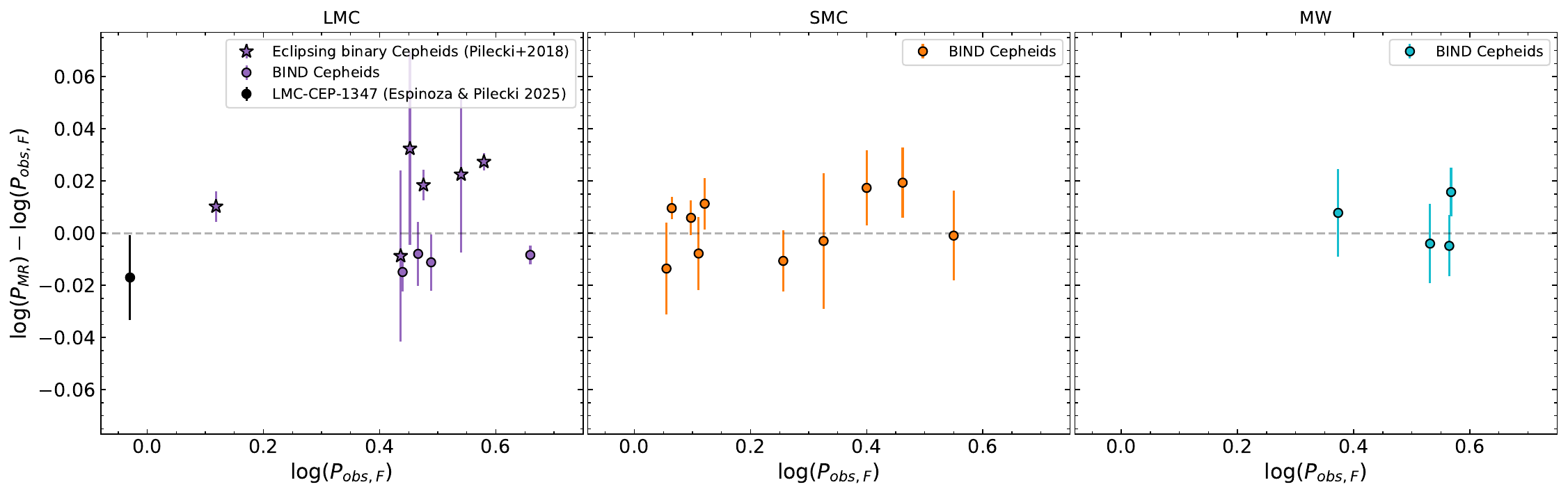}
    \caption{Same as Fig.\ref{fig:PMR-Pilecki} but for the PMR in eq.~\ref{eq:newPMR}.}
    \label{fig:PMR-New}
\end{figure*}

\end{appendix}

\end{document}